\documentclass[twocolumn,floats,floatfix,showpacs,amssymb,prd,superscriptaddress,nofootinbib]{revtex4-1}
\usepackage{graphicx, epsfig, amssymb} %include figure files
\usepackage{amsmath, amsfonts}
\usepackage{bm} 

\usepackage[linktocpage]{hyperref}
\usepackage[caption=false]{subfig}
\usepackage[usenames]{color}

\begin{document}

\title{\large Higher dimensional Numerical Relativity: code comparison}

%\title{\large Head-on collisions of BHs in $D=5$ and $D=6$ -- code comparison}

\author{Helvi Witek}\email{h.witek@damtp.cam.ac.uk}
\affiliation{Department of Applied Mathematics and Theoretical Physics,
Centre for Mathematical Sciences, University of Cambridge,
Wilberforce Road, Cambridge CB3 0WA, UK}

\author{Hirotada Okawa}\email{hirotada.okawa@ist.utl.pt}
\affiliation{CENTRA, Departamento de F\'{\i}sica, Instituto Superior T\'ecnico, Universidade de Lisboa,
Avenida Rovisco Pais 1, 1049 Lisboa, Portugal.}

\author{Vitor Cardoso}
\affiliation{CENTRA, Departamento de F\'{\i}sica, Instituto Superior T\'ecnico, Universidade de Lisboa,
Avenida Rovisco Pais 1, 1049 Lisboa, Portugal.}
\affiliation{Perimeter Institute for Theoretical Physics, Waterloo, Ontario N2L 2Y5, Canada.}
%\affiliation{Department of Physics and Astronomy, The University of Mississippi, University, MS 38677, USA.}

\author{Leonardo Gualtieri}
\affiliation{Dipartimento di Fisica, Universit\`a di Roma
``Sapienza'' \& Sezione INFN Roma1, P.A. Moro 5, 00185, Roma, Italy}

\author{Carlos Herdeiro}
\affiliation{Departamento de F\'\i sica da Universidade de Aveiro \& I3N, Campus de Santiago, 3810-183 Aveiro, Portugal}

\author{Masaru Shibata} %\email{mshibata@yukawa.kyoto-u.ac.jp}
\affiliation{Yukawa Institute for Theoretical Physics,
Kyoto University, 
Kyoto 606-8502, Japan}

\author{Ulrich Sperhake}
\affiliation{Department of Applied Mathematics and Theoretical Physics,
Centre for Mathematical Sciences, University of Cambridge,
Wilberforce Road, Cambridge CB3 0WA, UK}
\affiliation{Department of Physics and Astronomy, University of Mississippi,
University, Mississippi 38677, USA}
\affiliation{California Institute of Technology, Pasadena, California 91125, USA}

\author{Miguel Zilh\~ao}
\affiliation{Center for Computational Relativity and Gravitation and School of Mathematical Sciences,\\
Rochester Institute of Technology, Rochester, NY 14623, USA}

\begin{abstract}
The nonlinear behavior of higher dimensional black hole spacetimes is
of interest in several contexts, ranging from an understanding of cosmic censorship
to black hole production in high-energy collisions. However, nonlinear numerical evolutions of higher dimensional black hole spacetimes
are tremendously complex, involving different diagnostic tools and ``dimensional reduction methods''. 
In this work we compare two different successful codes to evolve Einstein's equations in higher dimensions, and show that
the results of such different procedures agree to numerical precision, when applied to the collision from rest of two equal-mass black holes.
We calculate the total radiated energy to be $E_{\rm rad}/M=(9.0\pm 0.5)\times 10^{-4}$ in five dimensions and $E_{\rm rad}/M=(8.1\pm 0.4)\times 10^{-4}$ in six dimensions.
\end{abstract}

%04.25.D-       Numerical relativity
%04.25.dg       Numerical studies of black holes and black-hole binaries
%04.50.-h       Higher-dimensional gravity and other theories of gravity
%04.50.Gh       Higher-dimensional black holes, black strings, and related objects

\maketitle

%\tableofcontents

%%%%%%%%%%%%%%%%%%%%%%%%%%%%%%%%%%%%%%%%%%%%%%%%%%%%%%%
\section{Introduction}\label{sec:Intro}
%%%%%%%%%%%%%%%%%%%%%%%%%%%%%%%%%%%%%%%%%%%%%%%%%%%%%%%

Higher-dimensional spacetimes have long played an important role
in theoretical physics. Such role has been highlighted in recent decades,
either through the realization of braneworld scenarios or in broader contexts
of quantum gravity theories, namely string theory. From a conceptual point of view, it is also useful --
and instructive -- to regard the spacetime dimensionality $D$ as one parameter more in the theory from which
to capitalize on to understand and gain intuition on the field equations.
The study of $D$-dimensional spacetimes has subsequently flourished, driven by many 
analytical or perturbative breakthroughs. A plethora of stationary black hole (BH) phases and their linear stability
properties have been studied~\cite{Emparan:2008eg,Emparan:2007wm,Harmark:2007md,Dias:2011jg,Dias:2014eua}. 
Simultaneously, exciting connections between dynamical black objects and the dynamics of fluids have been established~\cite{Cardoso:2006ks,Cardoso:2007ka,Bhattacharyya:2008jc,Emparan:2009at}.

Full-blown numerical methods are sometimes the only tool
to get an accurate, quantitative answer to  complex problems. It is a natural step in every exact science that
the resort to numerical methods becomes more frequent as the field matures.
Numerical Relativity -- the task of solving the dynamical gravitational field equations
in full generality -- has traditionally focused on four-dimensional, asymptotically flat spacetimes. The ``Holy Grail'' of the field was to solve and understand the two-body problem in General Relativity. Such attempts -- made successful in 2005 by several groups~\cite{Pretorius:2005gq,Campanelli:2005dd,Baker:2005vv} -- involve complex numerical
techniques and diagnostic tools, which had been developed during decades~\cite{LRR_group}. 
The intricacy of such problems and the need to calibrate -- and confirm -- results obtained
with some particular code, highlighted the need to compare different codes and results worldwide.
Such efforts have recently materialized for four-dimensional asymptotically flat spacetimes, in the context
of binary BHs as gravitational-wave (GW) sources~\cite{Aasi:2014tra,Hinder:2013oqa}.

Some of the numerical relativity results in higher dimensions are truly spectacular, and range
from black string fragmentation~\cite{Lehner:2010pn} to BH collisions~\cite{Witek:2010xi,Okawa:2011fv,Witek:2010az} and nonlinear instability growth~\cite{Shibata:2010wz} (for a review see Refs.~\cite{Yoshino:2011zz,Cardoso:2012qm,LRR_group}). These striking results, together with the potential 
of the field, call for a calibration of the different diagnostic tools and more urgently a comparison of different codes used to evolve
higher-dimensional spacetimes. A key purpose of this work is precisely to compare the two codes which have been developed to understand BH collisions and stability in higher dimensional spacetimes, namely the \textsc{HD-Lean}~\cite{Zilhao:2010sr,Witek:2010xi}
and the \textsc{SacraND} codes~\cite{Yoshino:2009xp,Okawa:2011fv}. In addition, we extend previous results to six-dimensional spacetimes.

%%%%%%%%%%%%%%%%%%%%%%%%%%%%%%%%%%%%%%%%%%%%%%%%%%%%%%%
\section{Numerical framework}\label{sec:NRframework}
%%%%%%%%%%%%%%%%%%%%%%%%%%%%%%%%%%%%%%%%%%%%%%%%%%%%%%%

Both codes, \textsc{SacraND} and \textsc{HD-Lean}, are based on
finite-differencing, ``3+1'' evolution schemes where Einstein's
equations are evolved
using the Baumgarte-Shapiro-Shibata-Nakamura (BSSN) formulation
\cite{Shibata:1995we,Baumgarte:1998te} combined with the
moving puncture method \cite{Campanelli:2005dd,Baker:2005vv};
for details of the respective 3+1 codes see
\cite{Yamamoto:2008js,Sperhake:2006cy}.
Higher, $D$-dimensional spacetimes with a $SO(D-3)$ [or $SO(D-2)$ for
the special case $D=5$] isometry
are accommodated in the
form of an effective 3+1 dimensional formulation with additional
fields that describe the extra dimensions, but the two codes differ
in the specific way in which this is achieved as well as in some
of the numerical technology and diagnostic tools.

\textsc{SacraND} uses the mesh-refinement algorithm described
in Ref.~\cite{Yamamoto:2008js}. 
%and extracts GWs using the
%Landau-Lifshitz pseudotensor; see for example Sec.~II in
%\cite{Lovelace:2009dg}. 
The Arnowitt-Deser-Misner
spacetime split \cite{Arnowitt:1962hi,York:1979} is applied to
the $D$-dimensional Einstein's equations which are translated
into a $D$-dimensional version of the BSSN equations. The spacetime
symmetry is then used to cast the equations into a 3+1 form
on a three-dimensional computational domain
with a modified version \cite{Shibata:2010wz,Yoshino:2011zz}
of the cartoon method originally introduced in \cite{Alcubierre:1999ab}.

\textsc{HD-Lean} is based on the {\sc Cactus} computational toolkit
\cite{Goodale:2002a,Cactuscode:web}, uses mesh refinement by
{\sc Carpet} \cite{Schnetter:2003rb} and {\sc AHFinderDirect}
\cite{Thornburg:1995cp,Thornburg:2003sf} for the calculation of
apparent horizons. In contrast to the {\sc SacraND} method, a
dimensional reduction is applied directly to the $D$ dimensional
Einstein equations analogous to Geroch's \cite{Geroch:1970nt}
decomposition; see also \cite{Cho:1986wk,Cho:1987jf}.
This results in the 3+1 Einstein equations
%non-minimally 
coupled to a scalar field which is converted into
a BSSN system with non-vanishing sources given by
the scalar field \cite{Zilhao:2010sr}. 

%%%%%%%%%%%%%%%%%%%%%%%%%%%%%%%%%%%%%%%%%%%%%%%%%%%%%%
\section{Wave extraction}
%%%%%%%%%%%%%%%%%%%%%%%%%%%%%%%%%%%%%%%%%%%%%%%%%%%%%%
Wave extraction is performed with two different approaches by the two codes.

The approach of \textsc{SacraND} (described in detail in \cite{Yoshino:2009xp})
is based on the fact that the spacetime is asymptotically flat. It is then possible
to describe the energy flux of the GWs produced in the
collision, in terms of the Landau-Lifshitz pseudo-tensor $t^{\mu\nu}_{LL}$
\cite{landau1975classical}, which has been generalized to a higher-dimensional
spacetime in Refs.~\cite{Cardoso:2002pa,Yoshino:2009xp}. The energy flux is
\begin{equation}
\frac{dE}{dt}=\int t_{LL}^{0i}n_idS\,,\label{Etll}
\end{equation}
where the integral is performed on a surface far away from the collision. We
remark that $t^{\mu\nu}_{LL}$ is not a tensor, but it behaves as a tensor
under general coordinate transformations of the background; in addition, the
total radiated energy obtained by integrating Eq.~(\ref{Etll}) is a gauge-invariant
quantity in the limit where the integration surface goes to inifinity.

The approach of {\sc HD-Lean} (described in detail in \cite{Witek:2010xi}) is
based on the fact that far away from the collision, the spacetime approaches a
spherically symmetric BH spacetime in higher dimensions i.e., the Tangherlini
BH solution \cite{Tangherlini:1963bw}. GWs produced in the
collision can be treated as perturbations of the Tangherlini solution using the
Kodama-Ishibashi (KI) formalism \cite{Kodama:2003jz,Ishibashi:2011ws}, which generalizes the
Regge-Wheeler-Zerilli-Moncrief formalism
\cite{Regge:1957td,Zerilli:1970se,Zerilli:1971wd,Moncrief:1974am} to higher
dimensions.

In the KI formalism the perturbations are expanded in tensor harmonics on the
$(D-2)$-sphere $S^{D-2}$. These harmonics belong to three classes: scalar, vector 
and tensor harmonics; the metric perturbations
associated to the different classes are decoupled in Einstein's equations. For each
of these classes, it is possible to define a (gauge-invariant) ``master
variable'' encoding the radiative degrees of freedom. Einstein's equations yield
a wave equation for each master variable.

As shown in \cite{Witek:2010xi}, in the case of head-on collisions the only
non-vanishing metric perturbations are those associated to scalar harmonics, due
to the $SO(D-3)$ isometry of the spacetime. The corresponding master variable
$\Phi^l$ (where $l\ge2$ is the index labelling the harmonic) can be constructed
in terms of the metric components, and it carries the GW energy
flux \cite{Berti:2003si}
\begin{equation}
\frac{dE_l}{dt}=\frac{1}{32\pi}\frac{D-3}{D-2}k^2(k^2-D+2)(\Phi^l_{,t})^2
\end{equation}
where $k^2=l(l+D-3)$.

%%%%%%%%%%%%%%%%%%%%%%%%%%%%%%%%%%%%%%%%%%%%%%%%%%%%%%%%%%%%%%%%%%%%%%
\section{Head-on collisions in $D=5$: comparison of results}\label{sec:Results5D}
%%%%%%%%%%%%%%%%%%%%%%%%%%%%%%%%%%%%%%%%%%%%%%%%%%%%%%%%%%%%%%%%%%%%%%
%%%%%%%%%%%%%%%%%%%%%%%%%%%%%%%%%%%%%%%%%%%%%%%%%%%%%%%%%%%%%%%%%%%%%%
\subsection{Runs}
%%%%%%%%%%%%%%%%%%%%%%%%%%%%%%%%%%%%%%%%%%%%%%%%%%%%%%%%%%%%%%%%%%%%%%
Black hole collisions in $D=5$ spacetime dimensions have been reported in the literature separately using both codes,
\textsc{HD-Lean}~\cite{Witek:2010xi} and \textsc{SacraND}~\cite{Okawa:2011fv}.
As such, we take $D=5$ to be the fiducial value for which to perform the comparison of results.
For this purpose we have performed a large set of simulations of equal-mass, non-rotating BH binaries
starting from rest with varying initial distance $d/r_{\rm{S}}=0.81,\ldots,12.93$ (where $r_{\rm S}$ is the Schwarzschild radius of the final BH).
The numerical domain typically consisted of $8$ nested grids where the two smallest refinement levels
contained components centered around each BH. For the results presented in this section
we typically used a (medium) resolution of $h/r_{\rm{S}}=1/84$ (for {\textsc{HD-Lean}}) 
and $h/r_{\rm{S}}=1/60$ (for {\textsc{SacraND}}) 
near the BHs resulting in, respectively, $h_{\rm{WE}}/r_{\rm{S}}=4/21$ and $h_{\rm{WE}}/r_{\rm{S}}=8/15$
in the wave zone.
 
%%%%%%%%%%%%%%%%%%%%%%%%%%%%%%%%%%%%%%%%%%%%%%%%%%%%%%%%%%%%%%%%%%%%%%
\subsection{Discretization and extrapolation error estimates}
%%%%%%%%%%%%%%%%%%%%%%%%%%%%%%%%%%%%%%%%%%%%%%%%%%%%%%%%%%%%%%%%%%%%%%
%
\begin{figure*}[htpb!]
 \subfloat[Convergence test (\textsc{SacraND})]{\includegraphics[width=0.52\textwidth,clip]{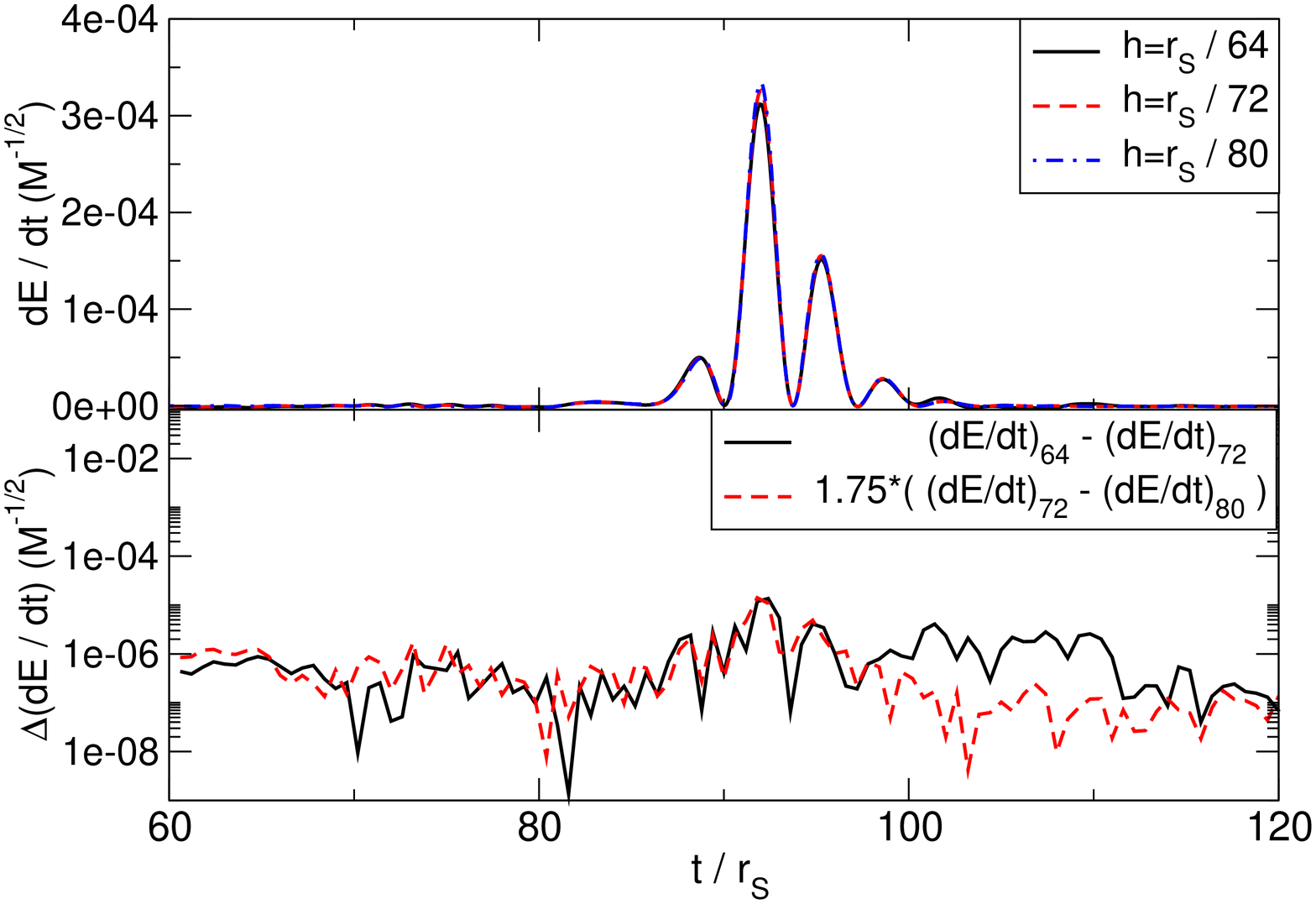}\label{fig:D5energyflux_K}}
 \subfloat[Convergence test (\textsc{HD-Lean})]{\includegraphics[width=0.52\textwidth,clip]{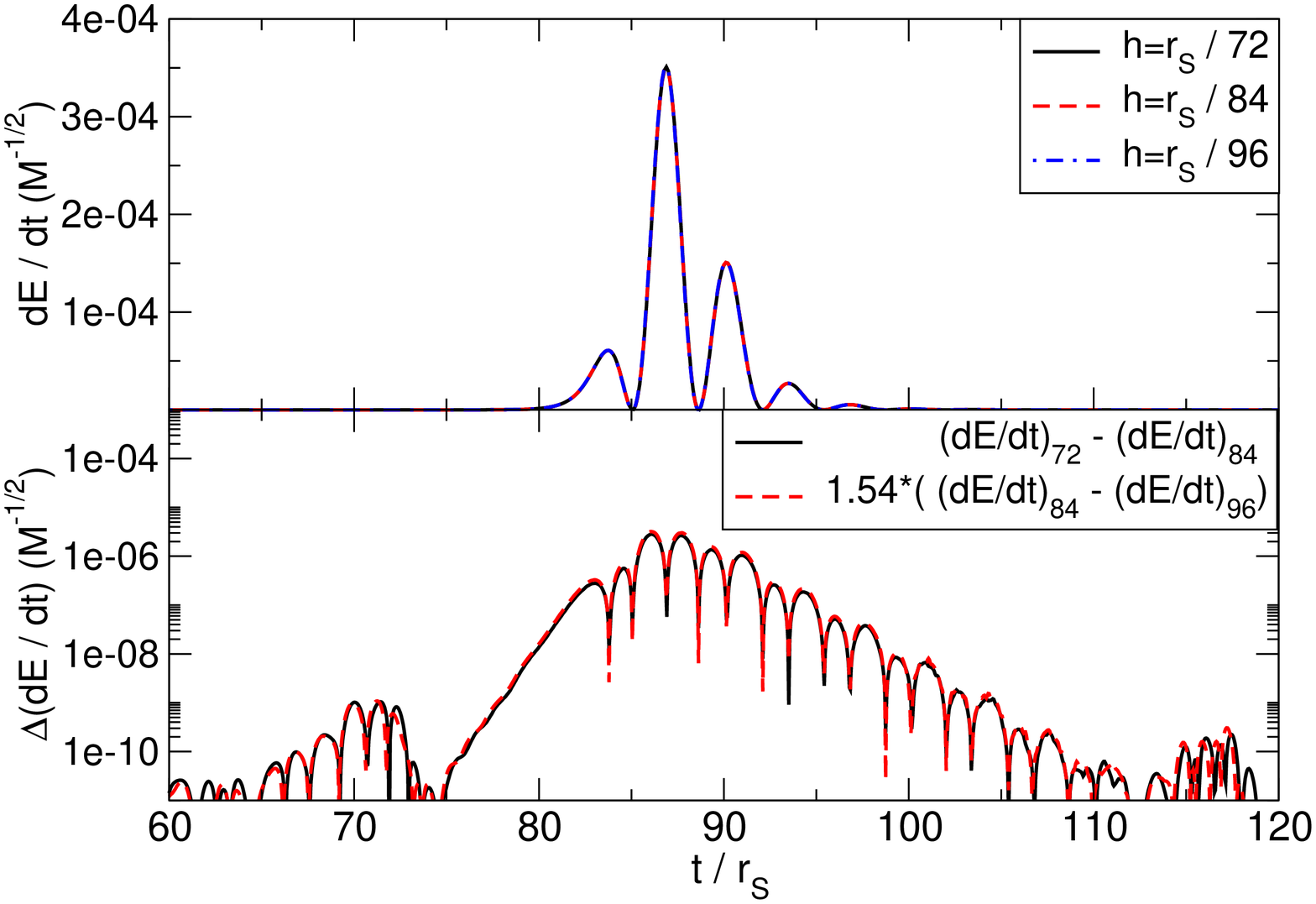}\label{fig:D5energyflux_L}}
 \caption{\label{fig:D5WaveformsConvergence}
 Convergence results for the head-on collision of two BHs in $D=5$ dimensions, for initial separation $d/r_{\rm{S}}=7$ and $d/r_{\rm{S}}=6.47$ 
(left and right panels respectively). Top panels refer to energy fluxes obtained respectively with \textsc{SacraND} and \textsc{HD-Lean} codes at different resolutions $h$. The corresponding convergence plots are shown in the bottom panels.
 Panel~\protect\subref{fig:D5energyflux_K} indicates 4th order convergence for the simulation performed with the \textsc{SacraND} code.
 Panel~\protect\subref{fig:D5energyflux_L} indicates 2nd order convergence for the simulations performed with the \textsc{HD-Lean} code.
}
\end{figure*}
A crucial component of our analysis involves the estimate of the numerical or discretization error which  
affects diagnostics using gravitational radiation output. 
In order to estimate the numerical accuracy of both codes we have performed convergence tests
using resolutions $h_{\rm{c}}/r_{\rm{S}}=1/72$, $h_{\rm{m}}/r_{\rm{S}}=1/84$ and $h_{\rm{h}}/r_{\rm{S}}=1/96$
(for {\textsc{HD-Lean}}) and $h_{\rm{c}}/r_{\rm{S}}=1/50$, $h_{\rm{m}}/r_{\rm{S}}=1/60$ and $h_{\rm{h}}/r_{\rm{S}}=3/200$ 
(for {\textsc{SacraND}}) near the BHs.

A convergence analysis, summarized in Fig.~\ref{fig:D5WaveformsConvergence} for the energy flux, shows second-order convergence for \textsc{HD-Lean} and fourth-order convergence for \textsc{SacraND}.
The numerical error in the total radiated energy is estimated to be
\begin{equation}
\Delta E_{\rm{rad}} / E_{\rm{rad}} \sim (0.7\%,\,1.3\%)
\end{equation}
for \textsc{HD-Lean} and \textsc{SacraND}, respectively.

%%%%%%%%%%%%%%%%%%%%%%%%%%%%%%%%%%%%%%%%%%%%%%%%%%%%%%%%%%%%%%%%%%%%%%
%\subsection{Extrapolation errors}
%%%%%%%%%%%%%%%%%%%%%%%%%%%%%%%%%%%%%%%%%%%%%%%%%%%%%%%%%%%%%%%%%%%%%%
In numerical time evolution codes, GW amplitudes are often measured at a finite ``extraction'' radius $r_{\rm ex}$ (but see
Refs.~\cite{Reisswig:2009us,Babiuc:2010ze} for exceptions developed for the four-dimensional case).
To compute physically relevant and unambiguous quantities, it is desirable
to extrapolate these quantities to $r_{\rm ex}/r_{\rm S} \to \infty$: fluxes and GW amplitudes are measured at a sphere of arbitrarily large radius.
The extrapolation procedure introduces an additional source of error.

We estimate the radiated energy as it would be measured at
$r_{\rm{ex}}/r_{\rm S}\rightarrow\infty$ by assuming an expansion of the form
\begin{align}
\label{eq:extrapolationD5}
E^{\rm{r_{ex}}}_{\rm{rad}}=& E^{\infty}_{\rm{rad}} + \sum_{j=1,2,...} A_j / r^{j}_{\rm{ex}}\,,
\end{align}
using the values $E^{r_{\rm{ex}}}_{\rm{rad}}$ calculated at fixed extraction radii to estimate
$E^{\infty}_{\rm{rad}}$ and the associated error.
We evaluate the error due to the extraction at finite radii and the extrapolation to be about $5\%$.

However, because our primary purpose here is to compare two different codes, we will compare their output
at finite $r_{\rm ex}$
and therefore  - since the quantities evaluated at finite $r_{\rm ex}$ are different in the two approaches - consider the $6~\%$ total uncertainty estimate
(from discretization and extrapolation)
a conservative upper limit for the expected discrepancies between the two codes at a given extraction radius.

%%%%%%%%%%%%%%%%%%%%%%%%%%%%%%%%%%%%%%%%%%%%%%%%%%%%%%%%%%%%%%%%%%%%%%
\subsection{(Spurious) Radiation content in initial data}
%%%%%%%%%%%%%%%%%%%%%%%%%%%%%%%%%%%%%%%%%%%%%%%%%%%%%%%%%%%%%%%%%%%%%%
Due to the initial data construction, in which we assume the maximal slicing condition $K=0$
and a finite initial distance between the BHs, the system contains a pulse of unphysical\footnote{``Unphysical'' in the sense that
a binary BH at rest at infinite initial separation would not be accompanied by such pulse of radiation at any finite distance} or spurious radiation, colloquially called ``junk'' radiation. This spurious radiation is typically emitted in a short burst after which the collision process proceeds normally. This is a well-known phenomenon in numerical relativity simulations
of BH binaries in $D=4$. Typically\footnote{But not always, for example high energy collisions evolving conformally flat initial data
are very challenging on account of the growing spurious radiation content at large Lorentz boosts~\cite{Sperhake:2008ga,Sperhake:2012me}.}, starting the collision at larger initial separations allows for the spurious radiation to leave the computational domain before interesting dynamics take place, therefore effectively eliminating its (undesirable) effect.

The energy content in the ``spurious'' pulse grows as the initial separation
between the two BHs decreases. This has been observed repeatedly in four-dimensional
spacetimes~\cite{Sperhake:2006cy,Lousto:2004pr}. We observe a similar pattern in $D=5$ spacetime dimensions.
For simulations with initial separation $d/r_{\rm S}=3,\,12$ for example, the fraction of the total radiated energy in the initial pulse is estimated to be $3\%$ and $<0.1\%$ respectively.
To avoid dealing with this spurious radiation we eliminate it from our analysis, by cutting it out of energy fluxes and waveforms;
this is possible for large enough initial separations, but becomes increasingly difficult for small initial separation.

%%%%%%%%%%%%%%%%%%%%%%%%%%%%%%%%%%%%%%%%%%%%%%%%%%%%%%%%%%%%%%%%%%%%%%
\subsection{Energy flux and total radiated energy}
%%%%%%%%%%%%%%%%%%%%%%%%%%%%%%%%%%%%%%%%%%%%%%%%%%%%%%%%%%%%%%%%%%%%%%
%
\begin{figure*}[htpb!]
 \subfloat[Energy fluxes]{\includegraphics[width=0.5\textwidth,clip]{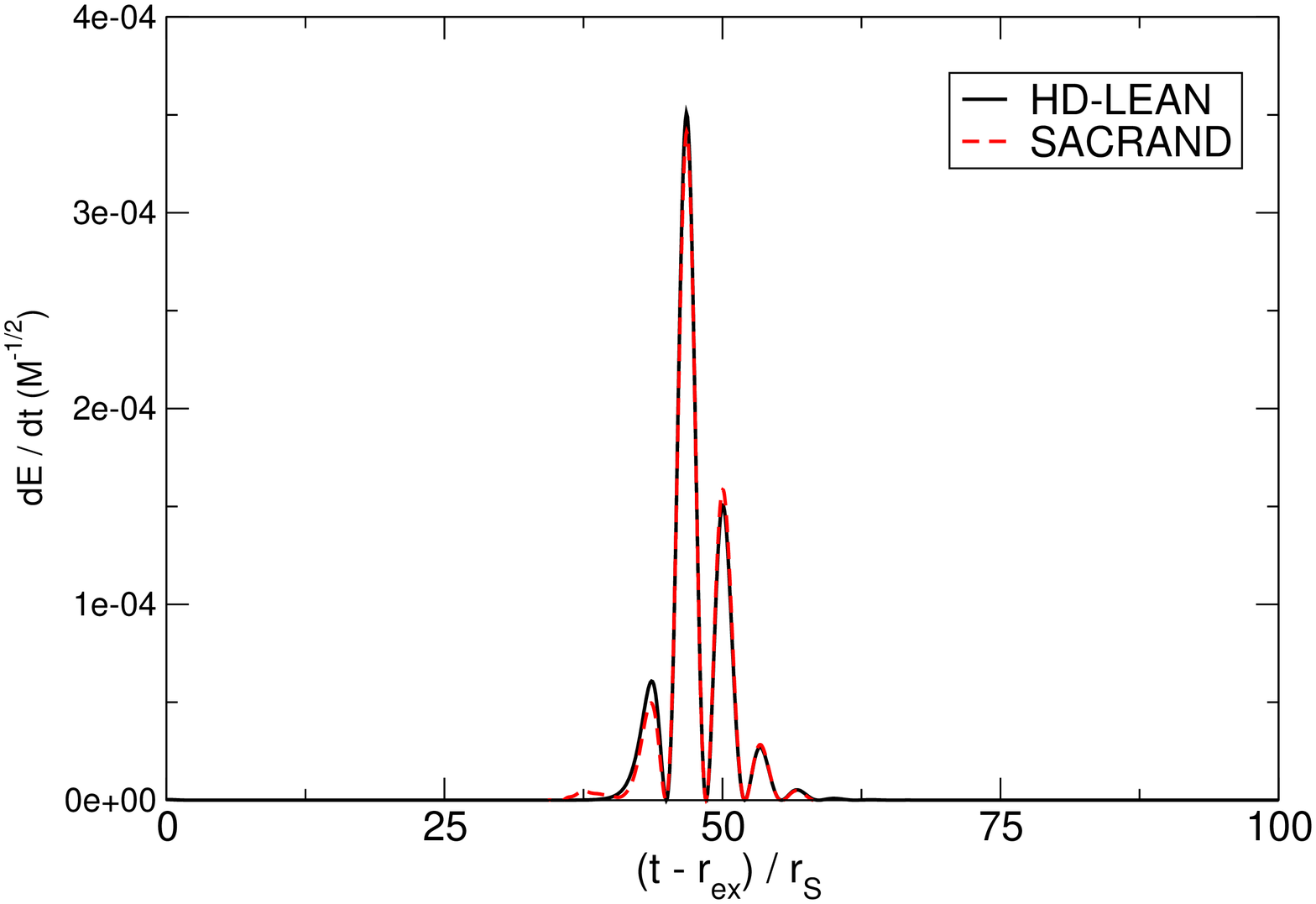}}
 \subfloat[Integrated fluxes]{\includegraphics[width=0.5\textwidth]{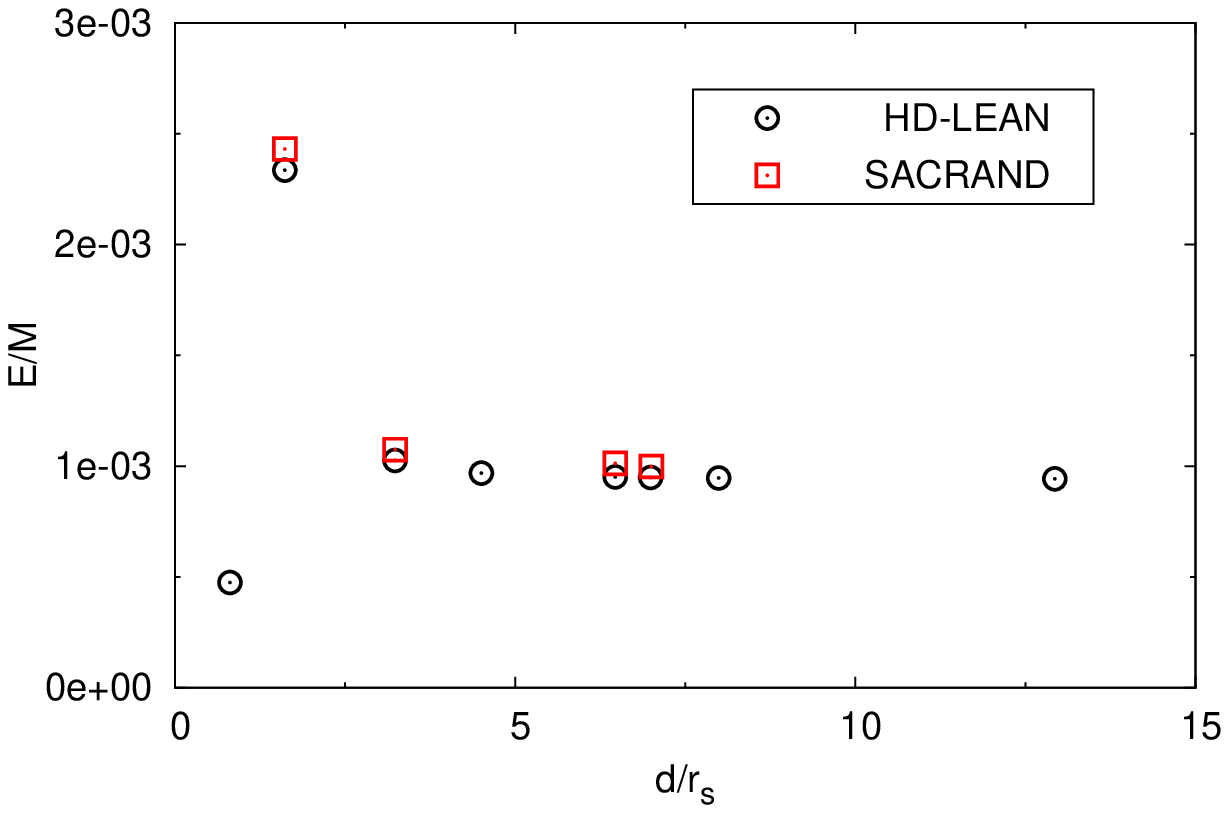}}
 \caption{\label{fig:D5energyflux_comparison}
 Energy fluxes for head-on collisions of two BHs in $D=5$ spacetime dimensions, obtained with \textsc{HD-Lean} (solid black line) and \textsc{SacraND} 
(red dashed line). The BHs start off at an initial coordinate separation $d/r_{\rm S}=6.47$. The right panel shows the total integrated energy for different BH initial separations.}
\end{figure*}
Results for the energy flux are shown in the left panel of Fig.~\ref{fig:D5energyflux_comparison}.
This is one of the main results of this work: both codes, using different dimensional reduction techniques
and different diagnostic tools, yield the same result for the flux and total radiated energy within numerical errors.

To avoid uncertainties with extrapolation methods, the energy flux shown in Fig.~\ref{fig:D5energyflux_comparison}
is measured at a finite coordinate radius of $r_{\rm ex}/r_{\rm S}=40$. The flux and waveforms show a clear ringdown
at late times. In particular, by fitting our numerical results to exponentially damped sinusoids, we estimate the $l=2$ and $l=4$ quasi-normal frequencies to be
\begin{subequations}
\label{eq:QNMNum5D}
\begin{align}
%D=5: & \quad
r_{\rm S}\,\omega_{l=2} = & 0.95 - \imath\, 0.26 \,,\\
r_{\rm S}\,\omega_{l=4} = & 2.12 - \imath\, 0.36\,.
\end{align}
\end{subequations}
We find good agreement with linearized predictions for the ringdown frequencies~\cite{Yoshino:2005ps,Berti:2009kk}
\begin{subequations}
\label{eq:QNMCL5D}
\begin{align}
%D=5: & \quad
r_{\rm S}\,\omega_{l=2} = & 0.9477 - \imath\, 0.2561 \,, \\
r_{\rm S}\,\omega_{l=4} = & 2.1924 - \imath\, 0.3293 
\,.
\end{align}
\end{subequations}

The right panel of Fig.~\ref{fig:D5energyflux_comparison} compares the total integrated energy for various initial separations using both codes.
The behavior with initial separation $d$ for five spacetime dimensions closely resembles the one found in four-dimensions~\cite{Anninos:1993zj,Lousto:1996sx}: at very small initial separations the binary closely resembles a single distorted BH,
and the radiation output is consequently very small. At large initial separations the radiation output asymptotes to a constant value,
\begin{equation}
E_{\rm rad}/M=(9.0\pm 0.5)\times 10^{-4}\label{eq:totD5}
\end{equation}
in agreement with Ref.~\cite{Witek:2010xi}.

There is a local maximum at finite initial separations, also reported in four-dimensional simulations in the point-particle limit~\cite{Lousto:1996sx}.
We highlight again the close agreement between the two codes. 
Finally, the area of the final AH allows one to estimate also the radiated energy. These are in good agreement with the estimates obtained via wave extraction~\cite{Witek:2010xi}.
%%%%%%%%%%%%%%%%%%%%%%%%%%%%%%%%%%%%%%%%%%%%%%%%%%%%%%%%%%%%%%%%%%%%%%%
\section{Head-on collisions in $D=6$}\label{sec:Results6D}
%%%%%%%%%%%%%%%%%%%%%%%%%%%%%%%%%%%%%%%%%%%%%%%%%%%%%%%%%%%%%%%%%%%%%%%
%
\begin{figure*}[htpb!]
\subfloat[Waveforms]{\includegraphics[width=0.5\textwidth,clip]{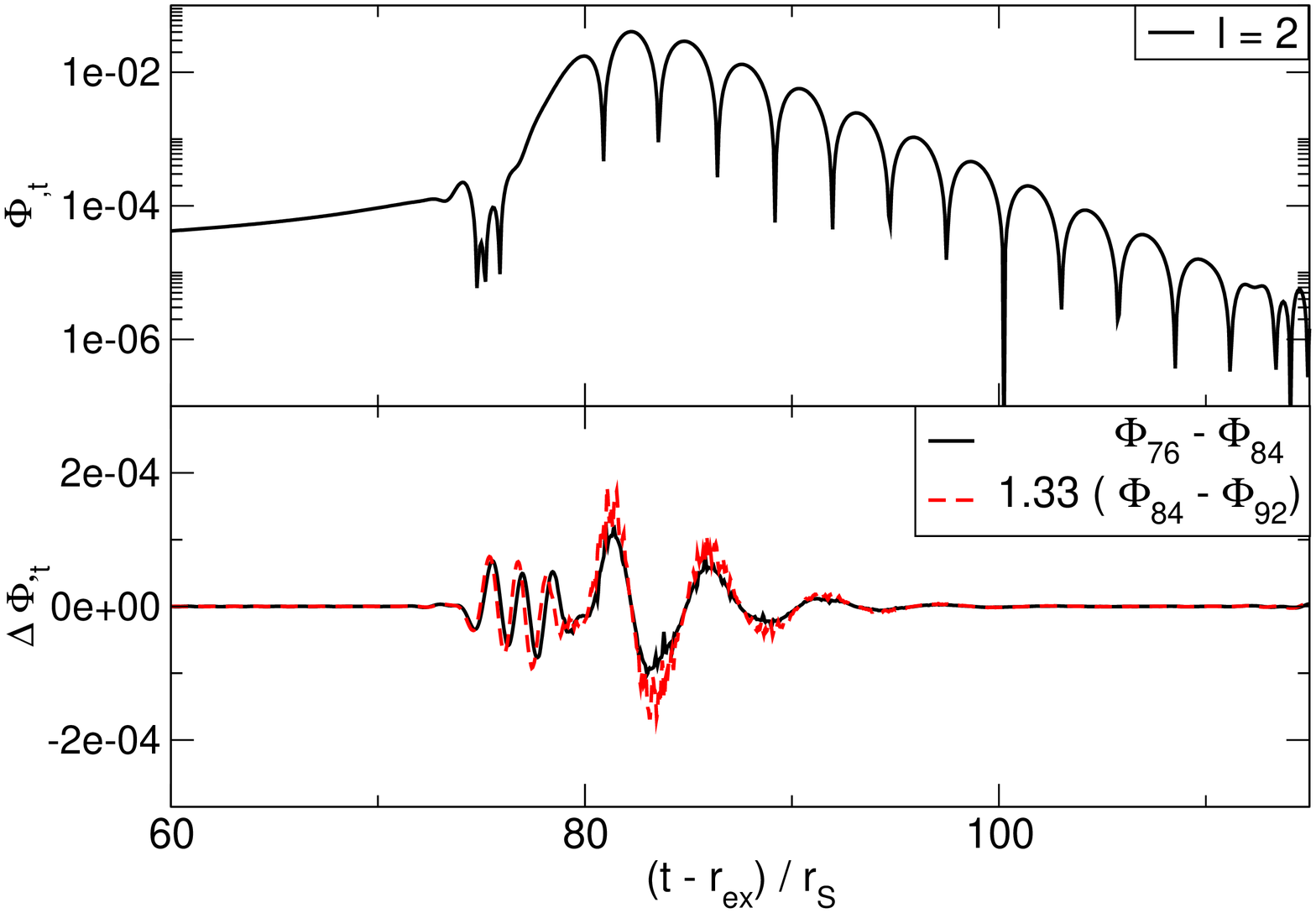}\label{fig:D6Waveforms}}
\subfloat[Energy]{\includegraphics[width=0.5\textwidth,clip]{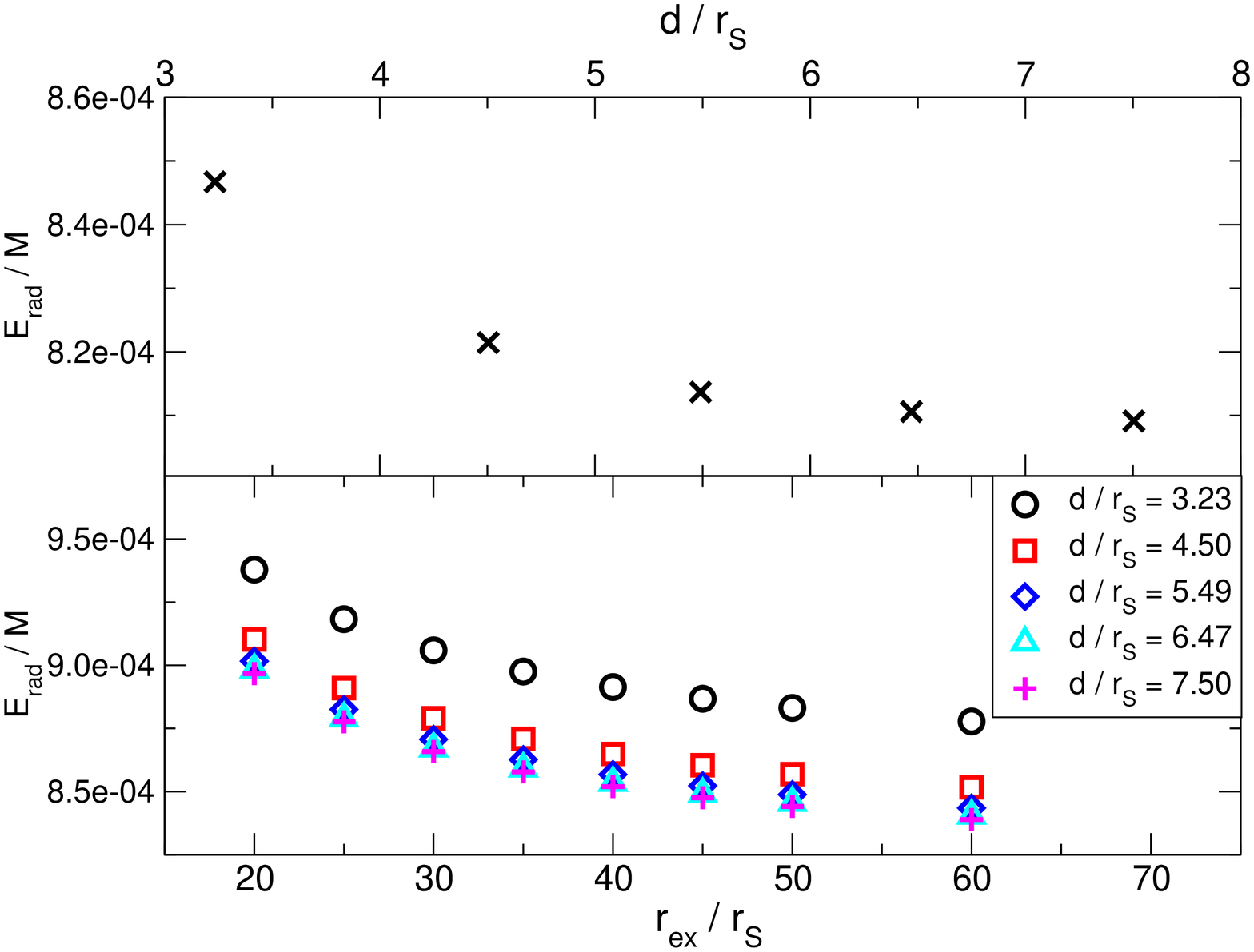}\label{fig:D6Energy}}
\caption{\label{fig:D6Plots} 
Summary of results for head-on collisions in $D=6$ spacetime dimensions.
Left: 
Time derivative of the $l=2$ multipole of the Kodama-Ishibashi gauge-invariant wavefunction $\Phi$ (top) 
and its convergence properties (bottom) computed with the \textsc{HD-Lean} code and
measured at the extraction radius $r_{\rm{ex}}/r_{\rm{S}}=40$.
The initial separation of the BHs has been $d/r_{\rm{S}}=6.47$.
The convergence factor $Q_{2}=1.33$ indicates second order convergence.
Right:
Total radiated energy $E_{\rm{rad}}/M$ emitted in the head-on collision of two BHs.  
The top panel shows the radiated energy, extrapolated to $r_{\rm{ex}}/r_{\rm{S}}\rightarrow\infty$
as a function of the initial separation $d/r_{\rm{S}}$.
The bottom panel shows the radiated energy as a function of the extraction radius 
$r_{\rm{ex}}/r_{\rm{S}}$ for different initial separations. 
}
\end{figure*}
Previous results in the literature concerning detailed analysis of BH collisions were specialized to four and five spacetime dimensions.
We now briefly describe results in $D=6$, summarized in Fig.~\ref{fig:D6Plots}.
A typical waveform is shown in the top left panel of Fig.~\ref{fig:D6Plots}. The GW signal, as measured by the gauge-invariant function $\Phi$,
displays the usual dominant quasi-normal ringdown. We estimate the ringdown parameters for the quadrupolar, $l=2$, component to be given by
$r_{\rm{S}}\,\omega_{l=2} = 1.14 - \imath\, 0.30$. This number compares very well against linearized calculations, which predict~\cite{Yoshino:2005ps,Berti:2009kk}
$r_{\rm{S}}\,\omega_{l=2} = 1.1369-\imath\,\,0.3038$.

As we mentioned previously, the computation of the energy flux is performed at a finite extraction radius. The total integrated flux yields the 
energy radiated in GWs and is consequently also computed at a finite location.
The physical total energy, computed at infinity, is estimated via extrapolation. These different quantities are shown in 
Fig.~\ref{fig:D6Plots} for different BH initial separations.

In the limit of infinite initial separation, the total radiated energy is
\begin{equation}
E_{\rm rad}/M=(8.1\pm 0.4)\times 10^{-4}\,.\label{eq:totD6}
\end{equation}
This number is comparable to, but smaller than the corresponding value in $D=5$ (see 
Eq.~\eqref{eq:totD5}), in agreement with the linearized point-particle calculations of Ref.~\cite{Berti:2010gx}.

%%%%%%%%%%%%%%%%%%%%%%%%%%%%%%%%%%%%%%%%%%%%%%%%%%%%%%%%%%%%%%%%%%%%%%%
\section{Conclusions}
%%%%%%%%%%%%%%%%%%%%%%%%%%%%%%%%%%%%%%%%%%%%%%%%%%%%%%%%%%%%%%%%%%%%%%%
Higher-dimensional spacetimes offer a vast and rich arena to test and understand the gravitational field equations.
The demand to understand, at a quantitative level, complex dynamical processes was met
by different, complex numerical codes and associated diagnostic tools.
The main purpose of this work is to show that the current numerical infrastructure
to handle BHs and BH-binaries in higher-dimensional spacetimes {\it is} solid and trustworthy.
We have compared two different Numerical Relativity codes, \textsc{HD-Lean}~\cite{Zilhao:2010sr,Witek:2010xi}
and \textsc{SacraND}~\cite{Yoshino:2009xp,Okawa:2011fv} which use different ``dimensional-reduction''
techniques and different wave extraction methods. Our main result is that both codes yield the same answer, up
to numerical errors which are under control. 
In addition, we determined the radiated energy in head-on collisions of six-dimensional black holes.

%%%%%%%%%%%%%%%%%%%%%%%%%%%%%%%%%%%%%%%%%%%%%%%%%%%%%%%
%\clearpage
%\newpage
\begin{acknowledgments}
%%%%%%%%%%%%%%%%%%%%%%%%%%%%%%%%%%%%%%%%%%%%%%%%%%%%%%%

We thank the Yukawa Institute for Theoretical Physics at
Kyoto University for hospitality during the YITP-T-14-1 
workshop on ``Holographic vistas on Gravity and Strings.''
H.~W. acknowledges financial support provided under
the {\it ERC-2011-StG 279363--HiDGR} ERC Starting Grant.
% and the STFC GR Roller grant ST/I002006/1.
%
H.O. and V.C. acknowledge financial support provided under the European Union's FP7 ERC Starting Grant ``The dynamics of black holes:
testing the limits of Einstein's theory'' grant agreement no. DyBHo--256667.
U.S. acknowledges support by
FP7-PEOPLE-2011-CIG Grant No. 293412 ``CBHEO" and
CESGA-ICTS Grant No. 249.
M.Z. is supported by NSF grants OCI-0832606, PHY-0969855, AST-1028087, and
PHY-1229173.
This research was supported in part by Perimeter Institute for Theoretical Physics. 
Research at Perimeter Institute is supported by the Government of Canada through 
Industry Canada and by the Province of Ontario through the Ministry of Economic Development 
$\&$ Innovation.
This work was supported by
the NRHEP 295189 FP7-PEOPLE-2011-IRSES Grant,
the STFC GR Roller Grant No. ST/L000636/1,
the Cosmos system, part of DiRAC, funded by STFC and BIS under
Grant Nos. ST/K00333X/1 and ST/J005673/1,
the NSF XSEDE Grant No. PHY-090003,
and by project PTDC/FIS/116625/2010.
Computations were performed on the ``Baltasar Sete-Sois'' cluster at IST,
on ``venus'' cluster at YITP,
on the COSMOS supercomputer in Cambridge,
the CESGA Finis Terrae, SDSC Trestles and NICS Kraken clusters.

\end{acknowledgments}

%%%%%%%%%%%%%%%%%%%%%%%%%%%%%%%%%%%%%%%%%%%%%%%%%%%%%%%
%\bibliographystyle{myutphys}
%\bibliographystyle{h-physrev4}
\bibliography{HDNumRel}

%merlin.mbs apsrev4-1.bst 2010-07-25 4.21a (PWD, AO, DPC) hacked
%Control: key (0)
%Control: author (8) initials jnrlst
%Control: editor formatted (1) identically to author
%Control: production of article title (-1) disabled
%Control: page (0) single
%Control: year (1) truncated
%Control: production of eprint (0) enabled
\begin{thebibliography}{59}%
\makeatletter
\providecommand \@ifxundefined [1]{%
 \@ifx{#1\undefined}
}%
\providecommand \@ifnum [1]{%
 \ifnum #1\expandafter \@firstoftwo
 \else \expandafter \@secondoftwo
 \fi
}%
\providecommand \@ifx [1]{%
 \ifx #1\expandafter \@firstoftwo
 \else \expandafter \@secondoftwo
 \fi
}%
\providecommand \natexlab [1]{#1}%
\providecommand \enquote  [1]{``#1''}%
\providecommand \bibnamefont  [1]{#1}%
\providecommand \bibfnamefont [1]{#1}%
\providecommand \citenamefont [1]{#1}%
\providecommand \href@noop [0]{\@secondoftwo}%
\providecommand \href [0]{\begingroup \@sanitize@url \@href}%
\providecommand \@href[1]{\@@startlink{#1}\@@href}%
\providecommand \@@href[1]{\endgroup#1\@@endlink}%
\providecommand \@sanitize@url [0]{\catcode `\\12\catcode `\$12\catcode
  `\&12\catcode `\#12\catcode `\^12\catcode `\_12\catcode `\%12\relax}%
\providecommand \@@startlink[1]{}%
\providecommand \@@endlink[0]{}%
\providecommand \url  [0]{\begingroup\@sanitize@url \@url }%
\providecommand \@url [1]{\endgroup\@href {#1}{\urlprefix }}%
\providecommand \urlprefix  [0]{URL }%
\providecommand \Eprint [0]{\href }%
\providecommand \doibase [0]{http://dx.doi.org/}%
\providecommand \selectlanguage [0]{\@gobble}%
\providecommand \bibinfo  [0]{\@secondoftwo}%
\providecommand \bibfield  [0]{\@secondoftwo}%
\providecommand \translation [1]{[#1]}%
\providecommand \BibitemOpen [0]{}%
\providecommand \bibitemStop [0]{}%
\providecommand \bibitemNoStop [0]{.\EOS\space}%
\providecommand \EOS [0]{\spacefactor3000\relax}%
\providecommand \BibitemShut  [1]{\csname bibitem#1\endcsname}%
\let\auto@bib@innerbib\@empty
%</preamble>
\bibitem [{\citenamefont {Emparan}\ and\ \citenamefont
  {Reall}(2008)}]{Emparan:2008eg}%
  \BibitemOpen
  \bibfield  {author} {\bibinfo {author} {\bibfnamefont {R.}~\bibnamefont
  {Emparan}}\ and\ \bibinfo {author} {\bibfnamefont {H.~S.}\ \bibnamefont
  {Reall}},\ }\href@noop {} {\bibfield  {journal} {\bibinfo  {journal} {Living
  Rev.Rel.}\ }\textbf {\bibinfo {volume} {11}},\ \bibinfo {pages} {6} (\bibinfo
  {year} {2008})},\ \Eprint {http://arxiv.org/abs/0801.3471} {arXiv:0801.3471
  [hep-th]} \BibitemShut {NoStop}%
%%CITATION = ARXIV:0801.3471;%%
\bibitem [{\citenamefont {Emparan}\ \emph {et~al.}(2007)\citenamefont
  {Emparan}, \citenamefont {Harmark}, \citenamefont {Niarchos}, \citenamefont
  {Obers},\ and\ \citenamefont {Rodriguez}}]{Emparan:2007wm}%
  \BibitemOpen
  \bibfield  {author} {\bibinfo {author} {\bibfnamefont {R.}~\bibnamefont
  {Emparan}}, \bibinfo {author} {\bibfnamefont {T.}~\bibnamefont {Harmark}},
  \bibinfo {author} {\bibfnamefont {V.}~\bibnamefont {Niarchos}}, \bibinfo
  {author} {\bibfnamefont {N.~A.}\ \bibnamefont {Obers}}, \ and\ \bibinfo
  {author} {\bibfnamefont {M.~J.}\ \bibnamefont {Rodriguez}},\ }\href {\doibase
  10.1088/1126-6708/2007/10/110} {\bibfield  {journal} {\bibinfo  {journal}
  {JHEP}\ }\textbf {\bibinfo {volume} {0710}},\ \bibinfo {pages} {110}
  (\bibinfo {year} {2007})},\ \Eprint {http://arxiv.org/abs/0708.2181}
  {arXiv:0708.2181 [hep-th]} \BibitemShut {NoStop}%
%%CITATION = ARXIV:0708.2181;%%
\bibitem [{\citenamefont {Harmark}\ \emph {et~al.}(2007)\citenamefont
  {Harmark}, \citenamefont {Niarchos},\ and\ \citenamefont
  {Obers}}]{Harmark:2007md}%
  \BibitemOpen
  \bibfield  {author} {\bibinfo {author} {\bibfnamefont {T.}~\bibnamefont
  {Harmark}}, \bibinfo {author} {\bibfnamefont {V.}~\bibnamefont {Niarchos}}, \
  and\ \bibinfo {author} {\bibfnamefont {N.~A.}\ \bibnamefont {Obers}},\ }\href
  {\doibase 10.1088/0264-9381/24/8/R01} {\bibfield  {journal} {\bibinfo
  {journal} {Class.Quant.Grav.}\ }\textbf {\bibinfo {volume} {24}},\ \bibinfo
  {pages} {R1} (\bibinfo {year} {2007})},\ \Eprint
  {http://arxiv.org/abs/hep-th/0701022} {arXiv:hep-th/0701022 [hep-th]}
  \BibitemShut {NoStop}%
%%CITATION = HEP-TH/0701022;%%
\bibitem [{\citenamefont {Dias}\ \emph {et~al.}(2011)\citenamefont {Dias},
  \citenamefont {Monteiro},\ and\ \citenamefont {Santos}}]{Dias:2011jg}%
  \BibitemOpen
  \bibfield  {author} {\bibinfo {author} {\bibfnamefont {O.~J.}\ \bibnamefont
  {Dias}}, \bibinfo {author} {\bibfnamefont {R.}~\bibnamefont {Monteiro}}, \
  and\ \bibinfo {author} {\bibfnamefont {J.~E.}\ \bibnamefont {Santos}},\
  }\href {\doibase 10.1007/JHEP08(2011)139} {\bibfield  {journal} {\bibinfo
  {journal} {JHEP}\ }\textbf {\bibinfo {volume} {1108}},\ \bibinfo {pages}
  {139} (\bibinfo {year} {2011})},\ \Eprint {http://arxiv.org/abs/1106.4554}
  {arXiv:1106.4554 [hep-th]} \BibitemShut {NoStop}%
%%CITATION = ARXIV:1106.4554;%%
\bibitem [{\citenamefont {Dias}\ \emph {et~al.}(2014)\citenamefont {Dias},
  \citenamefont {Hartnett},\ and\ \citenamefont {Santos}}]{Dias:2014eua}%
  \BibitemOpen
  \bibfield  {author} {\bibinfo {author} {\bibfnamefont {O.~J.~C.}\
  \bibnamefont {Dias}}, \bibinfo {author} {\bibfnamefont {G.~S.}\ \bibnamefont
  {Hartnett}}, \ and\ \bibinfo {author} {\bibfnamefont {J.~E.}\ \bibnamefont
  {Santos}},\ }\href@noop {} {\  (\bibinfo {year} {2014})},\ \Eprint
  {http://arxiv.org/abs/1402.7047} {arXiv:1402.7047 [hep-th]} \BibitemShut
  {NoStop}%
%%CITATION = ARXIV:1402.7047;%%
\bibitem [{\citenamefont {Cardoso}\ and\ \citenamefont
  {Dias}(2006)}]{Cardoso:2006ks}%
  \BibitemOpen
  \bibfield  {author} {\bibinfo {author} {\bibfnamefont {V.}~\bibnamefont
  {Cardoso}}\ and\ \bibinfo {author} {\bibfnamefont {O.~J.}\ \bibnamefont
  {Dias}},\ }\href {\doibase 10.1103/PhysRevLett.96.181601} {\bibfield
  {journal} {\bibinfo  {journal} {Phys.Rev.Lett.}\ }\textbf {\bibinfo {volume}
  {96}},\ \bibinfo {pages} {181601} (\bibinfo {year} {2006})},\ \Eprint
  {http://arxiv.org/abs/hep-th/0602017} {arXiv:hep-th/0602017 [hep-th]}
  \BibitemShut {NoStop}%
%%CITATION = HEP-TH/0602017;%%
\bibitem [{\citenamefont {Cardoso}\ \emph {et~al.}(2008)\citenamefont
  {Cardoso}, \citenamefont {Dias},\ and\ \citenamefont
  {Gualtieri}}]{Cardoso:2007ka}%
  \BibitemOpen
  \bibfield  {author} {\bibinfo {author} {\bibfnamefont {V.}~\bibnamefont
  {Cardoso}}, \bibinfo {author} {\bibfnamefont {O.~J.}\ \bibnamefont {Dias}}, \
  and\ \bibinfo {author} {\bibfnamefont {L.}~\bibnamefont {Gualtieri}},\ }\href
  {\doibase 10.1142/S0218271808012176} {\bibfield  {journal} {\bibinfo
  {journal} {Int.J.Mod.Phys.}\ }\textbf {\bibinfo {volume} {D17}},\ \bibinfo
  {pages} {505} (\bibinfo {year} {2008})},\ \Eprint
  {http://arxiv.org/abs/0705.2777} {arXiv:0705.2777 [hep-th]} \BibitemShut
  {NoStop}%
%%CITATION = ARXIV:0705.2777;%%
\bibitem [{\citenamefont {Bhattacharyya}\ \emph {et~al.}(2008)\citenamefont
  {Bhattacharyya}, \citenamefont {Hubeny}, \citenamefont {Minwalla},\ and\
  \citenamefont {Rangamani}}]{Bhattacharyya:2008jc}%
  \BibitemOpen
  \bibfield  {author} {\bibinfo {author} {\bibfnamefont {S.}~\bibnamefont
  {Bhattacharyya}}, \bibinfo {author} {\bibfnamefont {V.~E.}\ \bibnamefont
  {Hubeny}}, \bibinfo {author} {\bibfnamefont {S.}~\bibnamefont {Minwalla}}, \
  and\ \bibinfo {author} {\bibfnamefont {M.}~\bibnamefont {Rangamani}},\ }\href
  {\doibase 10.1088/1126-6708/2008/02/045} {\bibfield  {journal} {\bibinfo
  {journal} {JHEP}\ }\textbf {\bibinfo {volume} {0802}},\ \bibinfo {pages}
  {045} (\bibinfo {year} {2008})},\ \Eprint {http://arxiv.org/abs/0712.2456}
  {arXiv:0712.2456 [hep-th]} \BibitemShut {NoStop}%
%%CITATION = ARXIV:0712.2456;%%
\bibitem [{\citenamefont {Emparan}\ \emph {et~al.}(2010)\citenamefont
  {Emparan}, \citenamefont {Harmark}, \citenamefont {Niarchos},\ and\
  \citenamefont {Obers}}]{Emparan:2009at}%
  \BibitemOpen
  \bibfield  {author} {\bibinfo {author} {\bibfnamefont {R.}~\bibnamefont
  {Emparan}}, \bibinfo {author} {\bibfnamefont {T.}~\bibnamefont {Harmark}},
  \bibinfo {author} {\bibfnamefont {V.}~\bibnamefont {Niarchos}}, \ and\
  \bibinfo {author} {\bibfnamefont {N.~A.}\ \bibnamefont {Obers}},\ }\href
  {\doibase 10.1007/JHEP03(2010)063} {\bibfield  {journal} {\bibinfo  {journal}
  {JHEP}\ }\textbf {\bibinfo {volume} {1003}},\ \bibinfo {pages} {063}
  (\bibinfo {year} {2010})},\ \Eprint {http://arxiv.org/abs/0910.1601}
  {arXiv:0910.1601 [hep-th]} \BibitemShut {NoStop}%
%%CITATION = ARXIV:0910.1601;%%
\bibitem [{\citenamefont {Pretorius}(2005)}]{Pretorius:2005gq}%
  \BibitemOpen
  \bibfield  {author} {\bibinfo {author} {\bibfnamefont {F.}~\bibnamefont
  {Pretorius}},\ }\href {\doibase 10.1103/PhysRevLett.95.121101} {\bibfield
  {journal} {\bibinfo  {journal} {Phys.Rev.Lett.}\ }\textbf {\bibinfo {volume}
  {95}},\ \bibinfo {pages} {121101} (\bibinfo {year} {2005})},\ \Eprint
  {http://arxiv.org/abs/gr-qc/0507014} {arXiv:gr-qc/0507014 [gr-qc]}
  \BibitemShut {NoStop}%
%%CITATION = GR-QC/0507014;%%
\bibitem [{\citenamefont {Campanelli}\ \emph {et~al.}(2006)\citenamefont
  {Campanelli}, \citenamefont {Lousto}, \citenamefont {Marronetti},\ and\
  \citenamefont {Zlochower}}]{Campanelli:2005dd}%
  \BibitemOpen
  \bibfield  {author} {\bibinfo {author} {\bibfnamefont {M.}~\bibnamefont
  {Campanelli}}, \bibinfo {author} {\bibfnamefont {C.}~\bibnamefont {Lousto}},
  \bibinfo {author} {\bibfnamefont {P.}~\bibnamefont {Marronetti}}, \ and\
  \bibinfo {author} {\bibfnamefont {Y.}~\bibnamefont {Zlochower}},\ }\href
  {\doibase 10.1103/PhysRevLett.96.111101} {\bibfield  {journal} {\bibinfo
  {journal} {Phys.Rev.Lett.}\ }\textbf {\bibinfo {volume} {96}},\ \bibinfo
  {pages} {111101} (\bibinfo {year} {2006})},\ \Eprint
  {http://arxiv.org/abs/gr-qc/0511048} {arXiv:gr-qc/0511048 [gr-qc]}
  \BibitemShut {NoStop}%
%%CITATION = GR-QC/0511048;%%
\bibitem [{\citenamefont {Baker}\ \emph {et~al.}(2006)\citenamefont {Baker},
  \citenamefont {Centrella}, \citenamefont {Choi}, \citenamefont {Koppitz},\
  and\ \citenamefont {van Meter}}]{Baker:2005vv}%
  \BibitemOpen
  \bibfield  {author} {\bibinfo {author} {\bibfnamefont {J.~G.}\ \bibnamefont
  {Baker}}, \bibinfo {author} {\bibfnamefont {J.}~\bibnamefont {Centrella}},
  \bibinfo {author} {\bibfnamefont {D.-I.}\ \bibnamefont {Choi}}, \bibinfo
  {author} {\bibfnamefont {M.}~\bibnamefont {Koppitz}}, \ and\ \bibinfo
  {author} {\bibfnamefont {J.}~\bibnamefont {van Meter}},\ }\href {\doibase
  10.1103/PhysRevLett.96.111102} {\bibfield  {journal} {\bibinfo  {journal}
  {Phys.Rev.Lett.}\ }\textbf {\bibinfo {volume} {96}},\ \bibinfo {pages}
  {111102} (\bibinfo {year} {2006})},\ \Eprint
  {http://arxiv.org/abs/gr-qc/0511103} {arXiv:gr-qc/0511103 [gr-qc]}
  \BibitemShut {NoStop}%
%%CITATION = GR-QC/0511103;%%
\bibitem [{\citenamefont {Cardoso}\ \emph {et~al.}(2014)\citenamefont
  {Cardoso}, \citenamefont {Gualtieri}, \citenamefont {Herdeiro},\ and\
  \citenamefont {Sperhake}}]{LRR_group}%
  \BibitemOpen
  \bibfield  {author} {\bibinfo {author} {\bibfnamefont {V.}~\bibnamefont
  {Cardoso}}, \bibinfo {author} {\bibfnamefont {L.}~\bibnamefont {Gualtieri}},
  \bibinfo {author} {\bibfnamefont {C.}~\bibnamefont {Herdeiro}}, \ and\
  \bibinfo {author} {\bibfnamefont {U.}~\bibnamefont {Sperhake}},\ }\href@noop
  {} {\bibfield  {journal} {\bibinfo  {journal} {Living Reviews in Relativity
  (in preparation)}\ } (\bibinfo {year} {2014})}\BibitemShut {NoStop}%
\bibitem [{\citenamefont {Aasi}\ \emph {et~al.}(2014)\citenamefont {Aasi} \emph
  {et~al.}}]{Aasi:2014tra}%
  \BibitemOpen
  \bibfield  {author} {\bibinfo {author} {\bibfnamefont {J.}~\bibnamefont
  {Aasi}} \emph {et~al.} (\bibinfo {collaboration} {The LIGO Scientific
  Collaboration, the Virgo Collaboration, the NINJA-2 Collaboration}),\
  }\href@noop {} {\  (\bibinfo {year} {2014})},\ \Eprint
  {http://arxiv.org/abs/1401.0939} {arXiv:1401.0939 [gr-qc]} \BibitemShut
  {NoStop}%
%%CITATION = ARXIV:1401.0939;%%
\bibitem [{\citenamefont {Hinder}\ \emph {et~al.}(2014)\citenamefont {Hinder},
  \citenamefont {Buonanno}, \citenamefont {Boyle}, \citenamefont {Etienne},
  \citenamefont {Healy} \emph {et~al.}}]{Hinder:2013oqa}%
  \BibitemOpen
  \bibfield  {author} {\bibinfo {author} {\bibfnamefont {I.}~\bibnamefont
  {Hinder}}, \bibinfo {author} {\bibfnamefont {A.}~\bibnamefont {Buonanno}},
  \bibinfo {author} {\bibfnamefont {M.}~\bibnamefont {Boyle}}, \bibinfo
  {author} {\bibfnamefont {Z.~B.}\ \bibnamefont {Etienne}}, \bibinfo {author}
  {\bibfnamefont {J.}~\bibnamefont {Healy}},  \emph {et~al.},\ }\href {\doibase
  10.1088/0264-9381/31/2/025012} {\bibfield  {journal} {\bibinfo  {journal}
  {Class.Quant.Grav.}\ }\textbf {\bibinfo {volume} {31}},\ \bibinfo {pages}
  {025012} (\bibinfo {year} {2014})},\ \Eprint {http://arxiv.org/abs/1307.5307}
  {arXiv:1307.5307 [gr-qc]} \BibitemShut {NoStop}%
%%CITATION = ARXIV:1307.5307;%%
\bibitem [{\citenamefont {Lehner}\ and\ \citenamefont
  {Pretorius}(2010)}]{Lehner:2010pn}%
  \BibitemOpen
  \bibfield  {author} {\bibinfo {author} {\bibfnamefont {L.}~\bibnamefont
  {Lehner}}\ and\ \bibinfo {author} {\bibfnamefont {F.}~\bibnamefont
  {Pretorius}},\ }\href@noop {} {\bibfield  {journal} {\bibinfo  {journal}
  {Phys.Rev.Lett.}\ }\textbf {\bibinfo {volume} {105}},\ \bibinfo {pages}
  {101102} (\bibinfo {year} {2010})},\ \Eprint {http://arxiv.org/abs/1006.5960}
  {arXiv:1006.5960 [hep-th]} \BibitemShut {NoStop}%
%%CITATION = ARXIV:1006.5960;%%
\bibitem [{\citenamefont {Witek}\ \emph {et~al.}(2010)\citenamefont {Witek},
  \citenamefont {Zilhao}, \citenamefont {Gualtieri}, \citenamefont {Cardoso},
  \citenamefont {Herdeiro} \emph {et~al.}}]{Witek:2010xi}%
  \BibitemOpen
  \bibfield  {author} {\bibinfo {author} {\bibfnamefont {H.}~\bibnamefont
  {Witek}}, \bibinfo {author} {\bibfnamefont {M.}~\bibnamefont {Zilhao}},
  \bibinfo {author} {\bibfnamefont {L.}~\bibnamefont {Gualtieri}}, \bibinfo
  {author} {\bibfnamefont {V.}~\bibnamefont {Cardoso}}, \bibinfo {author}
  {\bibfnamefont {C.}~\bibnamefont {Herdeiro}},  \emph {et~al.},\ }\href
  {\doibase 10.1103/PhysRevD.82.104014} {\bibfield  {journal} {\bibinfo
  {journal} {Phys.Rev.}\ }\textbf {\bibinfo {volume} {D82}},\ \bibinfo {pages}
  {104014} (\bibinfo {year} {2010})},\ \Eprint {http://arxiv.org/abs/1006.3081}
  {arXiv:1006.3081 [gr-qc]} \BibitemShut {NoStop}%
%%CITATION = ARXIV:1006.3081;%%
\bibitem [{\citenamefont {Okawa}\ \emph {et~al.}(2011)\citenamefont {Okawa},
  \citenamefont {Nakao},\ and\ \citenamefont {Shibata}}]{Okawa:2011fv}%
  \BibitemOpen
  \bibfield  {author} {\bibinfo {author} {\bibfnamefont {H.}~\bibnamefont
  {Okawa}}, \bibinfo {author} {\bibfnamefont {K.-i.}\ \bibnamefont {Nakao}}, \
  and\ \bibinfo {author} {\bibfnamefont {M.}~\bibnamefont {Shibata}},\ }\href
  {\doibase 10.1103/PhysRevD.83.121501} {\bibfield  {journal} {\bibinfo
  {journal} {Phys.Rev.}\ }\textbf {\bibinfo {volume} {D83}},\ \bibinfo {pages}
  {121501} (\bibinfo {year} {2011})},\ \Eprint {http://arxiv.org/abs/1105.3331}
  {arXiv:1105.3331 [gr-qc]} \BibitemShut {NoStop}%
%%CITATION = ARXIV:1105.3331;%%
\bibitem [{\citenamefont {Witek}\ \emph {et~al.}(2011)\citenamefont {Witek},
  \citenamefont {Cardoso}, \citenamefont {Gualtieri}, \citenamefont {Herdeiro},
  \citenamefont {Sperhake} \emph {et~al.}}]{Witek:2010az}%
  \BibitemOpen
  \bibfield  {author} {\bibinfo {author} {\bibfnamefont {H.}~\bibnamefont
  {Witek}}, \bibinfo {author} {\bibfnamefont {V.}~\bibnamefont {Cardoso}},
  \bibinfo {author} {\bibfnamefont {L.}~\bibnamefont {Gualtieri}}, \bibinfo
  {author} {\bibfnamefont {C.}~\bibnamefont {Herdeiro}}, \bibinfo {author}
  {\bibfnamefont {U.}~\bibnamefont {Sperhake}},  \emph {et~al.},\ }\href
  {\doibase 10.1103/PhysRevD.83.044017} {\bibfield  {journal} {\bibinfo
  {journal} {Phys.Rev.}\ }\textbf {\bibinfo {volume} {D83}},\ \bibinfo {pages}
  {044017} (\bibinfo {year} {2011})},\ \Eprint {http://arxiv.org/abs/1011.0742}
  {arXiv:1011.0742 [gr-qc]} \BibitemShut {NoStop}%
%%CITATION = ARXIV:1011.0742;%%
\bibitem [{\citenamefont {Shibata}\ and\ \citenamefont
  {Yoshino}(2010)}]{Shibata:2010wz}%
  \BibitemOpen
  \bibfield  {author} {\bibinfo {author} {\bibfnamefont {M.}~\bibnamefont
  {Shibata}}\ and\ \bibinfo {author} {\bibfnamefont {H.}~\bibnamefont
  {Yoshino}},\ }\href {\doibase 10.1103/PhysRevD.81.104035} {\bibfield
  {journal} {\bibinfo  {journal} {Phys.Rev.}\ }\textbf {\bibinfo {volume}
  {D81}},\ \bibinfo {pages} {104035} (\bibinfo {year} {2010})},\ \Eprint
  {http://arxiv.org/abs/1004.4970} {arXiv:1004.4970 [gr-qc]} \BibitemShut
  {NoStop}%
%%CITATION = ARXIV:1004.4970;%%
\bibitem [{\citenamefont {Yoshino}\ and\ \citenamefont
  {Shibata}(2011)}]{Yoshino:2011zz}%
  \BibitemOpen
  \bibfield  {author} {\bibinfo {author} {\bibfnamefont {H.~M.~S.}\
  \bibnamefont {Yoshino}}\ and\ \bibinfo {author} {\bibfnamefont
  {M.}~\bibnamefont {Shibata}},\ }\href {\doibase 10.1143/PTPS.189.269}
  {\bibfield  {journal} {\bibinfo  {journal} {Prog.Theor.Phys.Suppl.}\ }\textbf
  {\bibinfo {volume} {189}},\ \bibinfo {pages} {269} (\bibinfo {year}
  {2011})}\BibitemShut {NoStop}%
%%CITATION = PTPSA,189,269;%%
\bibitem [{\citenamefont {Cardoso}\ \emph {et~al.}(2012)\citenamefont
  {Cardoso}, \citenamefont {Gualtieri}, \citenamefont {Herdeiro}, \citenamefont
  {Sperhake}, \citenamefont {Chesler} \emph {et~al.}}]{Cardoso:2012qm}%
  \BibitemOpen
  \bibfield  {author} {\bibinfo {author} {\bibfnamefont {V.}~\bibnamefont
  {Cardoso}}, \bibinfo {author} {\bibfnamefont {L.}~\bibnamefont {Gualtieri}},
  \bibinfo {author} {\bibfnamefont {C.}~\bibnamefont {Herdeiro}}, \bibinfo
  {author} {\bibfnamefont {U.}~\bibnamefont {Sperhake}}, \bibinfo {author}
  {\bibfnamefont {P.~M.}\ \bibnamefont {Chesler}},  \emph {et~al.},\ }\href
  {\doibase 10.1088/0264-9381/29/24/244001} {\bibfield  {journal} {\bibinfo
  {journal} {Class.Quant.Grav.}\ }\textbf {\bibinfo {volume} {29}},\ \bibinfo
  {pages} {244001} (\bibinfo {year} {2012})},\ \Eprint
  {http://arxiv.org/abs/1201.5118} {arXiv:1201.5118 [hep-th]} \BibitemShut
  {NoStop}%
%%CITATION = ARXIV:1201.5118;%%
\bibitem [{\citenamefont {Zilhao}\ \emph {et~al.}(2010)\citenamefont {Zilhao},
  \citenamefont {Witek}, \citenamefont {Sperhake}, \citenamefont {Cardoso},
  \citenamefont {Gualtieri} \emph {et~al.}}]{Zilhao:2010sr}%
  \BibitemOpen
  \bibfield  {author} {\bibinfo {author} {\bibfnamefont {M.}~\bibnamefont
  {Zilhao}}, \bibinfo {author} {\bibfnamefont {H.}~\bibnamefont {Witek}},
  \bibinfo {author} {\bibfnamefont {U.}~\bibnamefont {Sperhake}}, \bibinfo
  {author} {\bibfnamefont {V.}~\bibnamefont {Cardoso}}, \bibinfo {author}
  {\bibfnamefont {L.}~\bibnamefont {Gualtieri}},  \emph {et~al.},\ }\href
  {\doibase 10.1103/PhysRevD.81.084052} {\bibfield  {journal} {\bibinfo
  {journal} {Phys.Rev.}\ }\textbf {\bibinfo {volume} {D81}},\ \bibinfo {pages}
  {084052} (\bibinfo {year} {2010})},\ \Eprint {http://arxiv.org/abs/1001.2302}
  {arXiv:1001.2302 [gr-qc]} \BibitemShut {NoStop}%
%%CITATION = ARXIV:1001.2302;%%
\bibitem [{\citenamefont {Yoshino}\ and\ \citenamefont
  {Shibata}(2009)}]{Yoshino:2009xp}%
  \BibitemOpen
  \bibfield  {author} {\bibinfo {author} {\bibfnamefont {H.}~\bibnamefont
  {Yoshino}}\ and\ \bibinfo {author} {\bibfnamefont {M.}~\bibnamefont
  {Shibata}},\ }\href {\doibase 10.1103/PhysRevD.80.084025} {\bibfield
  {journal} {\bibinfo  {journal} {Phys.Rev.}\ }\textbf {\bibinfo {volume}
  {D80}},\ \bibinfo {pages} {084025} (\bibinfo {year} {2009})},\ \Eprint
  {http://arxiv.org/abs/0907.2760} {arXiv:0907.2760 [gr-qc]} \BibitemShut
  {NoStop}%
%%CITATION = ARXIV:0907.2760;%%
\bibitem [{\citenamefont {Shibata}\ and\ \citenamefont
  {Nakamura}(1995)}]{Shibata:1995we}%
  \BibitemOpen
  \bibfield  {author} {\bibinfo {author} {\bibfnamefont {M.}~\bibnamefont
  {Shibata}}\ and\ \bibinfo {author} {\bibfnamefont {T.}~\bibnamefont
  {Nakamura}},\ }\href {\doibase 10.1103/PhysRevD.52.5428} {\bibfield
  {journal} {\bibinfo  {journal} {Phys.Rev.}\ }\textbf {\bibinfo {volume}
  {D52}},\ \bibinfo {pages} {5428} (\bibinfo {year} {1995})}\BibitemShut
  {NoStop}%
%%CITATION = PHRVA,D52,5428;%%
\bibitem [{\citenamefont {Baumgarte}\ and\ \citenamefont
  {Shapiro}(1999)}]{Baumgarte:1998te}%
  \BibitemOpen
  \bibfield  {author} {\bibinfo {author} {\bibfnamefont {T.~W.}\ \bibnamefont
  {Baumgarte}}\ and\ \bibinfo {author} {\bibfnamefont {S.~L.}\ \bibnamefont
  {Shapiro}},\ }\href {\doibase 10.1103/PhysRevD.59.024007} {\bibfield
  {journal} {\bibinfo  {journal} {Phys.Rev.}\ }\textbf {\bibinfo {volume}
  {D59}},\ \bibinfo {pages} {024007} (\bibinfo {year} {1999})},\ \Eprint
  {http://arxiv.org/abs/gr-qc/9810065} {arXiv:gr-qc/9810065 [gr-qc]}
  \BibitemShut {NoStop}%
%%CITATION = GR-QC/9810065;%%
\bibitem [{\citenamefont {Yamamoto}\ \emph {et~al.}(2008)\citenamefont
  {Yamamoto}, \citenamefont {Shibata},\ and\ \citenamefont
  {Taniguchi}}]{Yamamoto:2008js}%
  \BibitemOpen
  \bibfield  {author} {\bibinfo {author} {\bibfnamefont {T.}~\bibnamefont
  {Yamamoto}}, \bibinfo {author} {\bibfnamefont {M.}~\bibnamefont {Shibata}}, \
  and\ \bibinfo {author} {\bibfnamefont {K.}~\bibnamefont {Taniguchi}},\ }\href
  {\doibase 10.1103/PhysRevD.78.064054} {\bibfield  {journal} {\bibinfo
  {journal} {Phys.Rev.}\ }\textbf {\bibinfo {volume} {D78}},\ \bibinfo {pages}
  {064054} (\bibinfo {year} {2008})},\ \Eprint {http://arxiv.org/abs/0806.4007}
  {arXiv:0806.4007 [gr-qc]} \BibitemShut {NoStop}%
%%CITATION = ARXIV:0806.4007;%%
\bibitem [{\citenamefont {Sperhake}(2007)}]{Sperhake:2006cy}%
  \BibitemOpen
  \bibfield  {author} {\bibinfo {author} {\bibfnamefont {U.}~\bibnamefont
  {Sperhake}},\ }\href {\doibase 10.1103/PhysRevD.76.104015} {\bibfield
  {journal} {\bibinfo  {journal} {Phys.Rev.}\ }\textbf {\bibinfo {volume}
  {D76}},\ \bibinfo {pages} {104015} (\bibinfo {year} {2007})},\ \Eprint
  {http://arxiv.org/abs/gr-qc/0606079} {arXiv:gr-qc/0606079 [gr-qc]}
  \BibitemShut {NoStop}%
%%CITATION = GR-QC/0606079;%%
\bibitem [{\citenamefont {Arnowitt}\ \emph {et~al.}(1962)\citenamefont
  {Arnowitt}, \citenamefont {Deser},\ and\ \citenamefont
  {Misner}}]{Arnowitt:1962hi}%
  \BibitemOpen
  \bibfield  {author} {\bibinfo {author} {\bibfnamefont {R.~L.}\ \bibnamefont
  {Arnowitt}}, \bibinfo {author} {\bibfnamefont {S.}~\bibnamefont {Deser}}, \
  and\ \bibinfo {author} {\bibfnamefont {C.~W.}\ \bibnamefont {Misner}},\
  }\href@noop {} {\  (\bibinfo {year} {1962})},\ \Eprint
  {http://arxiv.org/abs/gr-qc/0405109} {arXiv:gr-qc/0405109 [gr-qc]}
  \BibitemShut {NoStop}%
%%CITATION = GR-QC/0405109;%%
\bibitem [{\citenamefont {{York}}(1979)}]{York:1979}%
  \BibitemOpen
  \bibfield  {author} {\bibinfo {author} {\bibfnamefont {J.~W.}\ \bibnamefont
  {{York}}, \bibfnamefont {Jr.}},\ }in\ \href@noop {} {\emph {\bibinfo
  {booktitle} {Sources of Gravitational Radiation}}},\ \bibinfo {editor}
  {edited by\ \bibinfo {editor} {\bibnamefont {{L.~L.~Smarr}}}}\ (\bibinfo
  {year} {1979})\ pp.\ \bibinfo {pages} {83--126}\BibitemShut {NoStop}%
\bibitem [{\citenamefont {Alcubierre}\ \emph {et~al.}(2001)\citenamefont
  {Alcubierre}, \citenamefont {Brandt}, \citenamefont {Br{\"u}gmann},
  \citenamefont {Holz}, \citenamefont {Seidel}, \citenamefont {Takahashi},\
  and\ \citenamefont {Thornburg}}]{Alcubierre:1999ab}%
  \BibitemOpen
  \bibfield  {author} {\bibinfo {author} {\bibfnamefont {M.}~\bibnamefont
  {Alcubierre}}, \bibinfo {author} {\bibfnamefont {S.}~\bibnamefont {Brandt}},
  \bibinfo {author} {\bibfnamefont {B.}~\bibnamefont {Br{\"u}gmann}}, \bibinfo
  {author} {\bibfnamefont {D.}~\bibnamefont {Holz}}, \bibinfo {author}
  {\bibfnamefont {E.}~\bibnamefont {Seidel}}, \bibinfo {author} {\bibfnamefont
  {R.}~\bibnamefont {Takahashi}}, \ and\ \bibinfo {author} {\bibfnamefont
  {J.}~\bibnamefont {Thornburg}},\ }\href {\doibase 10.1142/S0218271801000834}
  {\bibfield  {journal} {\bibinfo  {journal} {Int. J. Mod. Phys. D}\ }\textbf
  {\bibinfo {volume} {10}},\ \bibinfo {pages} {273} (\bibinfo {year} {2001})},\
  \bibinfo {note} {gr-qc/9908012}\BibitemShut {NoStop}%
\bibitem [{\citenamefont {Goodale}\ \emph {et~al.}(2003)\citenamefont
  {Goodale}, \citenamefont {Allen}, \citenamefont {Lanfermann}, \citenamefont
  {Mass{\'o}}, \citenamefont {Radke}, \citenamefont {Seidel},\ and\
  \citenamefont {Shalf}}]{Goodale:2002a}%
  \BibitemOpen
  \bibfield  {author} {\bibinfo {author} {\bibfnamefont {T.}~\bibnamefont
  {Goodale}}, \bibinfo {author} {\bibfnamefont {G.}~\bibnamefont {Allen}},
  \bibinfo {author} {\bibfnamefont {G.}~\bibnamefont {Lanfermann}}, \bibinfo
  {author} {\bibfnamefont {J.}~\bibnamefont {Mass{\'o}}}, \bibinfo {author}
  {\bibfnamefont {T.}~\bibnamefont {Radke}}, \bibinfo {author} {\bibfnamefont
  {E.}~\bibnamefont {Seidel}}, \ and\ \bibinfo {author} {\bibfnamefont
  {J.}~\bibnamefont {Shalf}}\ }(\bibinfo  {publisher} {Springer},\ \bibinfo
  {address} {Berlin},\ \bibinfo {year} {2003})\BibitemShut {NoStop}%
\bibitem [{Cactus developers()}]{Cactuscode:web}%
  \BibitemOpen
  Cactus developers,\ \href {http://www.cactuscode.org/} {\enquote {\bibinfo
  {title} {{Cactus Computational Toolkit}},}\ }\BibitemShut {NoStop}%
\bibitem [{\citenamefont {Schnetter}\ \emph {et~al.}(2004)\citenamefont
  {Schnetter}, \citenamefont {Hawley},\ and\ \citenamefont
  {Hawke}}]{Schnetter:2003rb}%
  \BibitemOpen
  \bibfield  {author} {\bibinfo {author} {\bibfnamefont {E.}~\bibnamefont
  {Schnetter}}, \bibinfo {author} {\bibfnamefont {S.~H.}\ \bibnamefont
  {Hawley}}, \ and\ \bibinfo {author} {\bibfnamefont {I.}~\bibnamefont
  {Hawke}},\ }\href {\doibase 10.1088/0264-9381/21/6/014} {\bibfield  {journal}
  {\bibinfo  {journal} {Class.Quant.Grav.}\ }\textbf {\bibinfo {volume} {21}},\
  \bibinfo {pages} {1465} (\bibinfo {year} {2004})},\ \Eprint
  {http://arxiv.org/abs/gr-qc/0310042} {arXiv:gr-qc/0310042 [gr-qc]}
  \BibitemShut {NoStop}%
%%CITATION = GR-QC/0310042;%%
\bibitem [{\citenamefont {Thornburg}(1996)}]{Thornburg:1995cp}%
  \BibitemOpen
  \bibfield  {author} {\bibinfo {author} {\bibfnamefont {J.}~\bibnamefont
  {Thornburg}},\ }\href {\doibase 10.1103/PhysRevD.54.4899} {\bibfield
  {journal} {\bibinfo  {journal} {Phys.Rev.}\ }\textbf {\bibinfo {volume}
  {D54}},\ \bibinfo {pages} {4899} (\bibinfo {year} {1996})},\ \Eprint
  {http://arxiv.org/abs/gr-qc/9508014} {arXiv:gr-qc/9508014 [gr-qc]}
  \BibitemShut {NoStop}%
%%CITATION = GR-QC/9508014;%%
\bibitem [{\citenamefont {Thornburg}(2004)}]{Thornburg:2003sf}%
  \BibitemOpen
  \bibfield  {author} {\bibinfo {author} {\bibfnamefont {J.}~\bibnamefont
  {Thornburg}},\ }\href {\doibase 10.1088/0264-9381/21/2/026} {\bibfield
  {journal} {\bibinfo  {journal} {Class.Quant.Grav.}\ }\textbf {\bibinfo
  {volume} {21}},\ \bibinfo {pages} {743} (\bibinfo {year} {2004})},\ \Eprint
  {http://arxiv.org/abs/gr-qc/0306056} {arXiv:gr-qc/0306056 [gr-qc]}
  \BibitemShut {NoStop}%
%%CITATION = GR-QC/0306056;%%
\bibitem [{\citenamefont {Geroch}(1971)}]{Geroch:1970nt}%
  \BibitemOpen
  \bibfield  {author} {\bibinfo {author} {\bibfnamefont {R.~P.}\ \bibnamefont
  {Geroch}},\ }\href {\doibase 10.1063/1.1665681} {\bibfield  {journal}
  {\bibinfo  {journal} {J.Math.Phys.}\ }\textbf {\bibinfo {volume} {12}},\
  \bibinfo {pages} {918} (\bibinfo {year} {1971})}\BibitemShut {NoStop}%
%%CITATION = JMAPA,12,918;%%
\bibitem [{\citenamefont {Cho}(1987)}]{Cho:1986wk}%
  \BibitemOpen
  \bibfield  {author} {\bibinfo {author} {\bibfnamefont {Y.}~\bibnamefont
  {Cho}},\ }\href@noop {} {\bibfield  {journal} {\bibinfo  {journal}
  {Phys.Lett.}\ }\textbf {\bibinfo {volume} {186}},\ \bibinfo {pages} {38}
  (\bibinfo {year} {1987})}\BibitemShut {NoStop}%
%%CITATION = PHLTA,186,38;%%
\bibitem [{\citenamefont {Cho}\ and\ \citenamefont {Kim}(1989)}]{Cho:1987jf}%
  \BibitemOpen
  \bibfield  {author} {\bibinfo {author} {\bibfnamefont {Y.}~\bibnamefont
  {Cho}}\ and\ \bibinfo {author} {\bibfnamefont {D.}~\bibnamefont {Kim}},\
  }\href {\doibase 10.1063/1.528290} {\bibfield  {journal} {\bibinfo  {journal}
  {J.Math.Phys.}\ }\textbf {\bibinfo {volume} {30}},\ \bibinfo {pages} {1570}
  (\bibinfo {year} {1989})}\BibitemShut {NoStop}%
%%CITATION = JMAPA,30,1570;%%
\bibitem [{\citenamefont {Landau}\ and\ \citenamefont
  {Lifshitz}(1975)}]{landau1975classical}%
  \BibitemOpen
  \bibfield  {author} {\bibinfo {author} {\bibfnamefont {L.~D.}\ \bibnamefont
  {Landau}}\ and\ \bibinfo {author} {\bibfnamefont {E.~M.}\ \bibnamefont
  {Lifshitz}},\ }\href@noop {} {\emph {\bibinfo {title} {The classical theory
  of fields}}},\ Vol.~\bibinfo {volume} {2}\ (\bibinfo  {publisher}
  {Butterworth-Heinemann},\ \bibinfo {year} {1975})\BibitemShut {NoStop}%
\bibitem [{\citenamefont {Cardoso}\ \emph {et~al.}(2003)\citenamefont
  {Cardoso}, \citenamefont {Dias},\ and\ \citenamefont
  {Lemos}}]{Cardoso:2002pa}%
  \BibitemOpen
  \bibfield  {author} {\bibinfo {author} {\bibfnamefont {V.}~\bibnamefont
  {Cardoso}}, \bibinfo {author} {\bibfnamefont {O.~J.}\ \bibnamefont {Dias}}, \
  and\ \bibinfo {author} {\bibfnamefont {J.~P.}\ \bibnamefont {Lemos}},\ }\href
  {\doibase 10.1103/PhysRevD.67.064026} {\bibfield  {journal} {\bibinfo
  {journal} {Phys.Rev.}\ }\textbf {\bibinfo {volume} {D67}},\ \bibinfo {pages}
  {064026} (\bibinfo {year} {2003})},\ \Eprint
  {http://arxiv.org/abs/hep-th/0212168} {arXiv:hep-th/0212168 [hep-th]}
  \BibitemShut {NoStop}%
%%CITATION = HEP-TH/0212168;%%
\bibitem [{\citenamefont {Tangherlini}(1963)}]{Tangherlini:1963bw}%
  \BibitemOpen
  \bibfield  {author} {\bibinfo {author} {\bibfnamefont {F.}~\bibnamefont
  {Tangherlini}},\ }\href {\doibase 10.1007/BF02784569} {\bibfield  {journal}
  {\bibinfo  {journal} {Nuovo Cim.}\ }\textbf {\bibinfo {volume} {27}},\
  \bibinfo {pages} {636} (\bibinfo {year} {1963})}\BibitemShut {NoStop}%
%%CITATION = NUCIA,27,636;%%
\bibitem [{\citenamefont {Kodama}\ and\ \citenamefont
  {Ishibashi}(2003)}]{Kodama:2003jz}%
  \BibitemOpen
  \bibfield  {author} {\bibinfo {author} {\bibfnamefont {H.}~\bibnamefont
  {Kodama}}\ and\ \bibinfo {author} {\bibfnamefont {A.}~\bibnamefont
  {Ishibashi}},\ }\href {\doibase 10.1143/PTP.110.701} {\bibfield  {journal}
  {\bibinfo  {journal} {Prog.Theor.Phys.}\ }\textbf {\bibinfo {volume} {110}},\
  \bibinfo {pages} {701} (\bibinfo {year} {2003})},\ \Eprint
  {http://arxiv.org/abs/hep-th/0305147} {arXiv:hep-th/0305147 [hep-th]}
  \BibitemShut {NoStop}%
%%CITATION = HEP-TH/0305147;%%
\bibitem [{\citenamefont {Ishibashi}\ and\ \citenamefont
  {Kodama}(2011)}]{Ishibashi:2011ws}%
  \BibitemOpen
  \bibfield  {author} {\bibinfo {author} {\bibfnamefont {A.}~\bibnamefont
  {Ishibashi}}\ and\ \bibinfo {author} {\bibfnamefont {H.}~\bibnamefont
  {Kodama}},\ }\href {\doibase 10.1143/PTPS.189.165} {\bibfield  {journal}
  {\bibinfo  {journal} {Prog.Theor.Phys.Suppl.}\ }\textbf {\bibinfo {volume}
  {189}},\ \bibinfo {pages} {165} (\bibinfo {year} {2011})},\ \Eprint
  {http://arxiv.org/abs/1103.6148} {arXiv:1103.6148 [hep-th]} \BibitemShut
  {NoStop}%
%%CITATION = ARXIV:1103.6148;%%
\bibitem [{\citenamefont {Regge}\ and\ \citenamefont
  {Wheeler}(1957)}]{Regge:1957td}%
  \BibitemOpen
  \bibfield  {author} {\bibinfo {author} {\bibfnamefont {T.}~\bibnamefont
  {Regge}}\ and\ \bibinfo {author} {\bibfnamefont {J.~A.}\ \bibnamefont
  {Wheeler}},\ }\href {\doibase 10.1103/PhysRev.108.1063} {\bibfield  {journal}
  {\bibinfo  {journal} {Phys.Rev.}\ }\textbf {\bibinfo {volume} {108}},\
  \bibinfo {pages} {1063} (\bibinfo {year} {1957})}\BibitemShut {NoStop}%
%%CITATION = PHRVA,108,1063;%%
\bibitem [{\citenamefont {Zerilli}(1970{\natexlab{a}})}]{Zerilli:1970se}%
  \BibitemOpen
  \bibfield  {author} {\bibinfo {author} {\bibfnamefont {F.~J.}\ \bibnamefont
  {Zerilli}},\ }\href {\doibase 10.1103/PhysRevLett.24.737} {\bibfield
  {journal} {\bibinfo  {journal} {Phys.Rev.Lett.}\ }\textbf {\bibinfo {volume}
  {24}},\ \bibinfo {pages} {737} (\bibinfo {year}
  {1970}{\natexlab{a}})}\BibitemShut {NoStop}%
%%CITATION = PRLTA,24,737;%%
\bibitem [{\citenamefont {Zerilli}(1970{\natexlab{b}})}]{Zerilli:1971wd}%
  \BibitemOpen
  \bibfield  {author} {\bibinfo {author} {\bibfnamefont {F.~J.}\ \bibnamefont
  {Zerilli}},\ }\href {\doibase 10.1103/PhysRevD.2.2141} {\bibfield  {journal}
  {\bibinfo  {journal} {Phys. Rev. D}\ }\textbf {\bibinfo {volume} {2}},\
  \bibinfo {pages} {2141} (\bibinfo {year} {1970}{\natexlab{b}})}\BibitemShut
  {NoStop}%
\bibitem [{\citenamefont {Moncrief}(1974)}]{Moncrief:1974am}%
  \BibitemOpen
  \bibfield  {author} {\bibinfo {author} {\bibfnamefont {V.}~\bibnamefont
  {Moncrief}},\ }\href {\doibase 10.1016/0003-4916(74)90173-0} {\bibfield
  {journal} {\bibinfo  {journal} {Ann. Phys.}\ }\textbf {\bibinfo {volume}
  {88}},\ \bibinfo {pages} {323} (\bibinfo {year} {1974})}\BibitemShut
  {NoStop}%
\bibitem [{\citenamefont {Berti}\ \emph {et~al.}(2004)\citenamefont {Berti},
  \citenamefont {Cavaglia},\ and\ \citenamefont {Gualtieri}}]{Berti:2003si}%
  \BibitemOpen
  \bibfield  {author} {\bibinfo {author} {\bibfnamefont {E.}~\bibnamefont
  {Berti}}, \bibinfo {author} {\bibfnamefont {M.}~\bibnamefont {Cavaglia}}, \
  and\ \bibinfo {author} {\bibfnamefont {L.}~\bibnamefont {Gualtieri}},\ }\href
  {\doibase 10.1103/PhysRevD.69.124011} {\bibfield  {journal} {\bibinfo
  {journal} {Phys.Rev.}\ }\textbf {\bibinfo {volume} {D69}},\ \bibinfo {pages}
  {124011} (\bibinfo {year} {2004})},\ \Eprint
  {http://arxiv.org/abs/hep-th/0309203} {arXiv:hep-th/0309203 [hep-th]}
  \BibitemShut {NoStop}%
%%CITATION = HEP-TH/0309203;%%
\bibitem [{\citenamefont {Reisswig}\ \emph {et~al.}(2009)\citenamefont
  {Reisswig}, \citenamefont {Bishop}, \citenamefont {Pollney},\ and\
  \citenamefont {Szilagyi}}]{Reisswig:2009us}%
  \BibitemOpen
  \bibfield  {author} {\bibinfo {author} {\bibfnamefont {C.}~\bibnamefont
  {Reisswig}}, \bibinfo {author} {\bibfnamefont {N.}~\bibnamefont {Bishop}},
  \bibinfo {author} {\bibfnamefont {D.}~\bibnamefont {Pollney}}, \ and\
  \bibinfo {author} {\bibfnamefont {B.}~\bibnamefont {Szilagyi}},\ }\href
  {\doibase 10.1103/PhysRevLett.103.221101} {\bibfield  {journal} {\bibinfo
  {journal} {Phys.Rev.Lett.}\ }\textbf {\bibinfo {volume} {103}},\ \bibinfo
  {pages} {221101} (\bibinfo {year} {2009})},\ \Eprint
  {http://arxiv.org/abs/0907.2637} {arXiv:0907.2637 [gr-qc]} \BibitemShut
  {NoStop}%
%%CITATION = ARXIV:0907.2637;%%
\bibitem [{\citenamefont {Babiuc}\ \emph {et~al.}(2011)\citenamefont {Babiuc},
  \citenamefont {Szilagyi}, \citenamefont {Winicour},\ and\ \citenamefont
  {Zlochower}}]{Babiuc:2010ze}%
  \BibitemOpen
  \bibfield  {author} {\bibinfo {author} {\bibfnamefont {M.}~\bibnamefont
  {Babiuc}}, \bibinfo {author} {\bibfnamefont {B.}~\bibnamefont {Szilagyi}},
  \bibinfo {author} {\bibfnamefont {J.}~\bibnamefont {Winicour}}, \ and\
  \bibinfo {author} {\bibfnamefont {Y.}~\bibnamefont {Zlochower}},\ }\href
  {\doibase 10.1103/PhysRevD.84.044057} {\bibfield  {journal} {\bibinfo
  {journal} {Phys. Rev. D}\ }\textbf {\bibinfo {volume} {84}},\ \bibinfo
  {pages} {044057} (\bibinfo {year} {2011})},\ \bibinfo {note} {arXiv:1011.4223
  [gr-qc]}\BibitemShut {NoStop}%
\bibitem [{\citenamefont {Sperhake}\ \emph {et~al.}(2008)\citenamefont
  {Sperhake}, \citenamefont {Cardoso}, \citenamefont {Pretorius}, \citenamefont
  {Berti},\ and\ \citenamefont {Gonzalez}}]{Sperhake:2008ga}%
  \BibitemOpen
  \bibfield  {author} {\bibinfo {author} {\bibfnamefont {U.}~\bibnamefont
  {Sperhake}}, \bibinfo {author} {\bibfnamefont {V.}~\bibnamefont {Cardoso}},
  \bibinfo {author} {\bibfnamefont {F.}~\bibnamefont {Pretorius}}, \bibinfo
  {author} {\bibfnamefont {E.}~\bibnamefont {Berti}}, \ and\ \bibinfo {author}
  {\bibfnamefont {J.~A.}\ \bibnamefont {Gonzalez}},\ }\href {\doibase
  10.1103/PhysRevLett.101.161101} {\bibfield  {journal} {\bibinfo  {journal}
  {Phys.Rev.Lett.}\ }\textbf {\bibinfo {volume} {101}},\ \bibinfo {pages}
  {161101} (\bibinfo {year} {2008})},\ \Eprint {http://arxiv.org/abs/0806.1738}
  {arXiv:0806.1738 [gr-qc]} \BibitemShut {NoStop}%
%%CITATION = ARXIV:0806.1738;%%
\bibitem [{\citenamefont {Sperhake}\ \emph {et~al.}(2013)\citenamefont
  {Sperhake}, \citenamefont {Berti}, \citenamefont {Cardoso},\ and\
  \citenamefont {Pretorius}}]{Sperhake:2012me}%
  \BibitemOpen
  \bibfield  {author} {\bibinfo {author} {\bibfnamefont {U.}~\bibnamefont
  {Sperhake}}, \bibinfo {author} {\bibfnamefont {E.}~\bibnamefont {Berti}},
  \bibinfo {author} {\bibfnamefont {V.}~\bibnamefont {Cardoso}}, \ and\
  \bibinfo {author} {\bibfnamefont {F.}~\bibnamefont {Pretorius}},\ }\href
  {\doibase 10.1103/PhysRevLett.111.041101} {\bibfield  {journal} {\bibinfo
  {journal} {Phys. Rev. Lett.}\ }\textbf {\bibinfo {volume} {111}},\ \bibinfo
  {pages} {041101} (\bibinfo {year} {2013})},\ \bibinfo {note} {arXiv:1211.6114
  [gr-qc]}\BibitemShut {NoStop}%
\bibitem [{\citenamefont {Lousto}\ and\ \citenamefont
  {Price}(2004)}]{Lousto:2004pr}%
  \BibitemOpen
  \bibfield  {author} {\bibinfo {author} {\bibfnamefont {C.}~\bibnamefont
  {Lousto}}\ and\ \bibinfo {author} {\bibfnamefont {R.~H.}\ \bibnamefont
  {Price}},\ }\href {\doibase 10.1103/PhysRevD.69.087503} {\bibfield  {journal}
  {\bibinfo  {journal} {Phys.Rev.}\ }\textbf {\bibinfo {volume} {D69}},\
  \bibinfo {pages} {087503} (\bibinfo {year} {2004})},\ \Eprint
  {http://arxiv.org/abs/gr-qc/0401045} {arXiv:gr-qc/0401045 [gr-qc]}
  \BibitemShut {NoStop}%
%%CITATION = GR-QC/0401045;%%
\bibitem [{\citenamefont {Yoshino}\ \emph {et~al.}(2005)\citenamefont
  {Yoshino}, \citenamefont {Shiromizu},\ and\ \citenamefont
  {Shibata}}]{Yoshino:2005ps}%
  \BibitemOpen
  \bibfield  {author} {\bibinfo {author} {\bibfnamefont {H.}~\bibnamefont
  {Yoshino}}, \bibinfo {author} {\bibfnamefont {T.}~\bibnamefont {Shiromizu}},
  \ and\ \bibinfo {author} {\bibfnamefont {M.}~\bibnamefont {Shibata}},\ }\href
  {\doibase 10.1103/PhysRevD.72.084020} {\bibfield  {journal} {\bibinfo
  {journal} {Phys.Rev.}\ }\textbf {\bibinfo {volume} {D72}},\ \bibinfo {pages}
  {084020} (\bibinfo {year} {2005})},\ \Eprint
  {http://arxiv.org/abs/gr-qc/0508063} {arXiv:gr-qc/0508063 [gr-qc]}
  \BibitemShut {NoStop}%
%%CITATION = GR-QC/0508063;%%
\bibitem [{\citenamefont {Berti}\ \emph {et~al.}(2009)\citenamefont {Berti},
  \citenamefont {Cardoso},\ and\ \citenamefont {Starinets}}]{Berti:2009kk}%
  \BibitemOpen
  \bibfield  {author} {\bibinfo {author} {\bibfnamefont {E.}~\bibnamefont
  {Berti}}, \bibinfo {author} {\bibfnamefont {V.}~\bibnamefont {Cardoso}}, \
  and\ \bibinfo {author} {\bibfnamefont {A.~O.}\ \bibnamefont {Starinets}},\
  }\href {\doibase 10.1088/0264-9381/26/16/163001} {\bibfield  {journal}
  {\bibinfo  {journal} {Class.Quant.Grav.}\ }\textbf {\bibinfo {volume} {26}},\
  \bibinfo {pages} {163001} (\bibinfo {year} {2009})},\ \Eprint
  {http://arxiv.org/abs/0905.2975} {arXiv:0905.2975 [gr-qc]} \BibitemShut
  {NoStop}%
%%CITATION = ARXIV:0905.2975;%%
\bibitem [{\citenamefont {Anninos}\ \emph {et~al.}(1993)\citenamefont
  {Anninos}, \citenamefont {Hobill}, \citenamefont {Seidel}, \citenamefont
  {Smarr},\ and\ \citenamefont {Suen}}]{Anninos:1993zj}%
  \BibitemOpen
  \bibfield  {author} {\bibinfo {author} {\bibfnamefont {P.}~\bibnamefont
  {Anninos}}, \bibinfo {author} {\bibfnamefont {D.}~\bibnamefont {Hobill}},
  \bibinfo {author} {\bibfnamefont {E.}~\bibnamefont {Seidel}}, \bibinfo
  {author} {\bibfnamefont {L.}~\bibnamefont {Smarr}}, \ and\ \bibinfo {author}
  {\bibfnamefont {W.-M.}\ \bibnamefont {Suen}},\ }\href {\doibase
  10.1103/PhysRevLett.71.2851} {\bibfield  {journal} {\bibinfo  {journal}
  {Phys.Rev.Lett.}\ }\textbf {\bibinfo {volume} {71}},\ \bibinfo {pages} {2851}
  (\bibinfo {year} {1993})},\ \Eprint {http://arxiv.org/abs/gr-qc/9309016}
  {arXiv:gr-qc/9309016 [gr-qc]} \BibitemShut {NoStop}%
%%CITATION = GR-QC/9309016;%%
\bibitem [{\citenamefont {Lousto}\ and\ \citenamefont
  {Price}(1997)}]{Lousto:1996sx}%
  \BibitemOpen
  \bibfield  {author} {\bibinfo {author} {\bibfnamefont {C.~O.}\ \bibnamefont
  {Lousto}}\ and\ \bibinfo {author} {\bibfnamefont {R.~H.}\ \bibnamefont
  {Price}},\ }\href {\doibase 10.1103/PhysRevD.55.2124} {\bibfield  {journal}
  {\bibinfo  {journal} {Phys.Rev.}\ }\textbf {\bibinfo {volume} {D55}},\
  \bibinfo {pages} {2124} (\bibinfo {year} {1997})},\ \Eprint
  {http://arxiv.org/abs/gr-qc/9609012} {arXiv:gr-qc/9609012 [gr-qc]}
  \BibitemShut {NoStop}%
%%CITATION = GR-QC/9609012;%%
\bibitem [{\citenamefont {Berti}\ \emph {et~al.}(2011)\citenamefont {Berti},
  \citenamefont {Cardoso},\ and\ \citenamefont {Kipapa}}]{Berti:2010gx}%
  \BibitemOpen
  \bibfield  {author} {\bibinfo {author} {\bibfnamefont {E.}~\bibnamefont
  {Berti}}, \bibinfo {author} {\bibfnamefont {V.}~\bibnamefont {Cardoso}}, \
  and\ \bibinfo {author} {\bibfnamefont {B.}~\bibnamefont {Kipapa}},\ }\href
  {\doibase 10.1103/PhysRevD.83.084018} {\bibfield  {journal} {\bibinfo
  {journal} {Phys.Rev.}\ }\textbf {\bibinfo {volume} {D83}},\ \bibinfo {pages}
  {084018} (\bibinfo {year} {2011})},\ \Eprint {http://arxiv.org/abs/1010.3874}
  {arXiv:1010.3874 [gr-qc]} \BibitemShut {NoStop}%
%%CITATION = ARXIV:1010.3874;%%
\end{thebibliography}%

\end{document}